\documentclass[journal=jacsat,manuscript=article]{achemso}

\usepackage{chemformula} 
\usepackage[T1]{fontenc} 
\usepackage[version=3]{mhchem} 
\usepackage[font=small,labelfont=bf]{caption}
\mciteErrorOnUnknownfalse
\usepackage{geometry}
\usepackage[english]{babel}
\usepackage[utf8x]{inputenc}
\usepackage[T1]{fontenc}
\usepackage{helvet}  
\usepackage{comment}
\usepackage{color}

\newcommand{\tcb}{\textcolor{blue}}

\newcommand{\SI}{\tcb{SI}}

\newcommand{\dm}{ \Delta \mu}
\newcommand{\Tm}{T_{\rm eq}}
\newcommand{\dt}{ \Delta T}
\newcommand{\dts}{ \Delta T^* }
\newcommand{\Gbarr}{G_{\rm barr}}
\newcommand{\glam}{g_\lambda}
\newcommand{\gm}{g_{\rm m}}
\newcommand{\mg}{m_g}
\newcommand{\cg}{c_g}
\newcommand{\Gint}{G_{\rm th}}
\newcommand{\dgam}{\Delta\gamma}
\newcommand{\lagmult}{\mathcal{L}_\lambda}
\newcommand{\hstar}{\bar{h}_\lambda}

\newcommand{\lamb}{\lambda_{\rm b}}
\newcommand{\lamm}{\lambda_{\rm m}}
\newcommand{\kaplam}{\kappa_\lambda}
\newcommand{\kapm}{\kappa_{\rm m}}
\newcommand{\intf}{\mathcal{I}}
\newcommand{\conl}{\partial\mathcal{I}}
\newcommand{\nhat}{\hat{n}}
\newcommand{\mhat}{\hat{m}}

\newcommand{\am}{a_{\rm m}}
\newcommand{\Pm}{p_{\rm m}}
\newcommand{\etam}{\eta_{\rm m}}
\newcommand{\Lsq}{L^2}
\newcommand{\Lsqinv}{L^{-2}}

\newcommand{\Linv}{L^{-1}}

\author{Aniket U. Thosar}
\author{Yusheng Cai}
\author{Sean M. Marks}
\author{Zachariah Vicars}
\author{Jeongmoon Choi}
\author{Akash Pallath}
\author{Amish J. Patel}
\email{amish.patel@seas.upenn.edu}
\affiliation{Department of Chemical and Biomolecular Engineering, University of Pennsylvania}

\title{On the Engulfment of Antifreeze Proteins by Ice}

\begin{document}

\newpage

\begin{abstract}
Antifreeze proteins (AFPs) are remarkable biomolecules that suppress ice formation at trace concentrations. 
To inhibit ice growth, AFPs must not only bind to ice crystals, but also resist engulfment by ice.
The highest supercooling, $\dts$, for which AFPs are able to resist engulfment is widely believed to scale as the inverse of the separation, $L$, between bound AFPs, whereas its dependence on the molecular characteristics of the AFP remains poorly understood.
By using specialized molecular simulations and interfacial thermodynamics, 
here we 
show that in contrast with conventional wisdom, $\dts$ scales as $\Lsqinv$ and not as $\Linv$.
We further show that $\dts$ is proportional to AFP size and that diverse naturally occurring AFPs are optimal at resisting engulfment by ice.
By facilitating the development of AFP structure-function relationships, 
we hope that our findings will pave the way for the rational design of novel AFPs. 
\end{abstract}

Antifreeze proteins (AFPs) facilitate the survival of diverse organisms in frigid environments by suppressing ice formation in their cells and bodily fluids~\cite{devries1969adsorption,AFPs_from_diverse_organisms_paper_1,haji2016rating}.
Remarkably, only trace AFP concentrations (roughly $\mu$M) are necessary to inhibit the growth of ice crystals~\cite{AFPs_from_diverse_organisms_paper_2,tyshenko1997antifreeze}.
AFPs 
are also 
potent inhibitors of ice recrystallization~\cite{knight1984a, budke2014quantitative}.
Consequently, AFPs are being explored for a variety of applications, ranging from the transportation and storage of frozen food to the cryopreservation of transplant organs~\cite{AFP_application_I, murray2022chemical, wang2018antifreeze}.
Several creative studies have shown that to inhibit the growth of ice crystals,
AFPs have two sides with very different but equally important functions~\cite{raymond1977adsorption, Knight:2001db, kristiansen2005mechanism, drori2015dts_L_dependance, naullage2018controls, bianco2020antifreeze}: 
the ice-binding side (IBS) of an AFP facilitates its binding to ice, 
whereas its non-binding side (NBS) enables it to resist engulfment by ice~\cite{AFP_ice_binding, AFP_ice_binding_Ran_paper_PNAS, Molinero_JACS_II, meister2015investigation, Molinero_PNAS, marks2018antifreeze}.

%
The ability of bound AFPs to resist engulfment depends on the supercooling, $\dt \equiv \Tm - T$, where $\Tm$ is the equilibrium melting temperature and $T$ is the system temperature.
Although most AFPs are able to resist engulfment at low $\dt$, 
at sufficiently high $\dt$, all AFPs are eventually engulfed by ice~\cite{kristiansen2005mechanism,naullage2018controls}.
The highest supercooling for which an AFP resists engulfment, $\dts$, thus informs the temperature range over which ice growth is suppressed, i.e., the thermal hysteresis activity of the AFP~\cite{naullage2018controls,graham1997a}.
Both experiments and molecular simulations have shown that $\dts$ depends not only on the type of AFP, but also on the separation, $L$, between bound AFPs~\cite{drori2015dts_L_dependance, bianco2020antifreeze}.
In particular, inspired by the Gibbs-Thompson relationship,
which relates the melting temperature of ice crystals to their size,
it has been suggested that $\dts \sim \Linv$~\cite{raymond1977adsorption, kristiansen2005mechanism}.
However, whether AFP engulfment should obey such a scaling~\cite{sander2004kinetic, farag2023free, bianco2020antifreeze} 
and importantly, how the molecular characteristics of the NBS influence $\dts$ remain open questions~\cite{pal2022molecular, zanetti-polzi2019a, karlsson2018protein}.
%

To address these questions, here we employ a combination of molecular simulations, enhanced sampling techniques~\cite{xi2018sparse,marks2023characterizing} and interfacial thermodynamics~\cite{gennes2004capillarity}. 
We characterize the free energetics of AFP engulfment by ice and find that 
although engulfment is impeded by large barriers at low $\dt$, 
those barriers decrease as $\dt$ is increased, and vanish altogether at $\dts$.
Using interfacial thermodynamics, we then elucidate how engulfment barriers 
stem from a competition between the driving force for ice growth and 
the ice-phobicity of the NBS, 
which results in the pinning of the ice-water interface to the NBS and a curving of the interface.
We also show that $\dts$ is proportional to the highest interfacial curvature 
that can be sustained by bound AFPs.
Importantly, in contrast with the widely accepted scaling, $\dts \sim \Linv$~\cite{raymond1977adsorption, Knight:2001db, kristiansen2005mechanism, nada2012antifreeze_review_paper_MDsim, drori2015dts_L_dependance, bianco2020antifreeze, kuiper2015biological, midya2018operation, kozuch2018combined, naullage2018controls, gerhauser2021detailed}, 
we find that both interfacial curvature and $\dts$ scale as $\Lsqinv$ and are proportional to AFP size.
We also quantify the effectiveness of AFPs at resisting engulfment,
and find that the NBS of diverse naturally-occurring AFPs are optimally suited for this task.
Finally, we find that the ability of an AFP to resist engulfment is relatively insensitive to the presence of charged residues on its NBS.

\begin{figure*}[htbp]
\centering
\includegraphics[width=1.\textwidth]{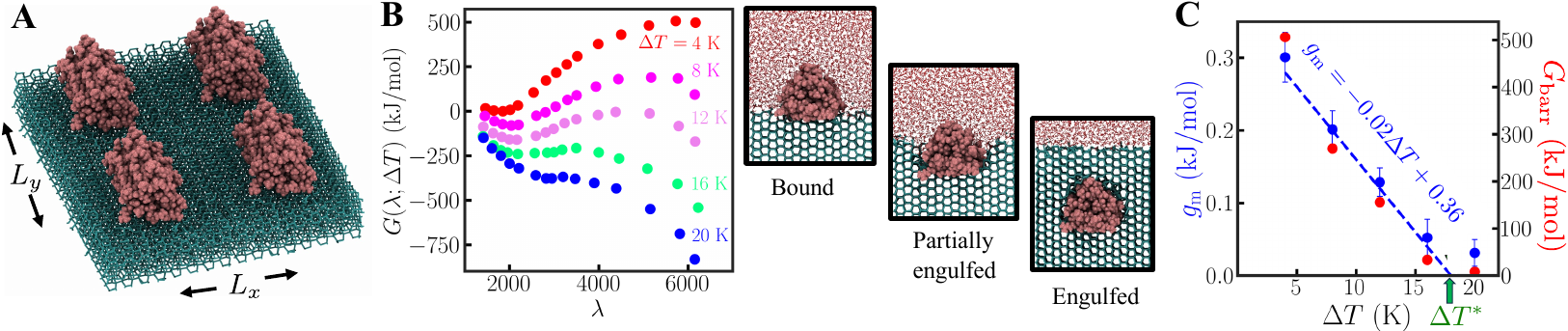}
\caption{Free energetics of antifreeze protein (AFP) engulfment by ice. 
(A) Simulation snapshots of the spruce budworm AFP (sbwAFP) bound to the ice-water interface; the AFPs are shown in pink, ice is in cyan, and liquid waters are hidden for clarity.
Due to periodic boundary conditions, the AFPs are arranged in a rectangular array with separations, $L_x$ and $L_y$, of roughly 6~nm along the $x$- and $y$-dimensions, respectively.
(B) At low values of supercooling, $\dt$, the free energetics of engulfment, $G(\lambda; \dt)$, 
display large barriers, which arrest ice growth; here, $\lambda$ is the number of ice-like waters.
As $\dt$ is increased, the free energetic barriers decrease and eventually vanish above $\dts \approx 18$~K, leading to the spontaneous engulfment of the AFP.
Representative configurations with the AFP bound to ice as well as with the AFP partially and fully engulfed by ice are shown with liquid waters in red.
(C) The barrier to engulfment, $\Gbarr$ (right), correlates well with $\gm$ (left), the maximum of the slope, $\glam \equiv dG/d\lambda$; 
both quantities decrease with increasing supercooling, $\dt$ (points), and vanish at $\dts$. 
The variation of $\gm$ with $\dt$ is captured well by a linear fit, $\gm(\dt) = - \mg \dt + \cg$ (dashed line), so the supercooling at which $\gm$ vanishes can also be obtained as $\dts= \cg/\mg \approx 18$~K.
}
\label{fig:Schematic}
\end{figure*}

\section*{Free Energetics of AFP Engulfment by Ice}
To understand how AFPs resist engulfment by ice,
we first study the spruce budworm AFP (sbwAFP)~\cite{tyshenko1997antifreeze, leinala2002beta}.
For hyperactive AFPs, such as sbwAFP, which bind to all ice planes~\cite{scotter2006basis}, 
AFP engulfment results in rapid and uncontrolled ice growth~\cite{Basu20012015}.
Thus, the supercooling, $\dts$, at which a hyperactive AFP is spontaneously engulfed by ice, 
also provides a quantitative measure of its thermal hysteresis activity~\cite{drori2014ice_dts_conc_annealing, naullage2018controls}.
To characterize sbwAFP engulfment by ice and its dependence on supercooling, $\dt$, 
we estimate the free energetics, $G(\lambda; \dt)$, of observing $\lambda$ ice-like waters 
in the system
for a wide range of $\dt$-values from 4~K to 20~K.
A simulation snapshot of sbwAFP molecules bound to the ice-water interface, and separated by roughly 6~nm, is shown in Figure~\ref{fig:Schematic}A, and
a detailed description of the simulation setup and parameters as well as the enhanced sampling techniques used to estimate $G(\lambda; \dt)$ is provided in the \SI.
%

As shown in Figure~\ref{fig:Schematic}B, $G(\lambda; \dt)$ displays a local minimum 
at $\lambda \approx 2000$ corresponding to the AFP bound to ice.
For low supercooling, the bound (low $\lambda$) and engulfed (high $\lambda$) states are separated by large free energetic barriers, $\Gbarr$, which arrest ice growth and localize the system in its bound state. 
However, as $\dt$ is increased, and the thermodynamic driving force for ice growth increases, the engulfment barriers decrease (Figure~\ref{fig:Schematic}B). 
Above a critical supercooling, $\dts \approx 18$~K, the engulfment barrier, $\Gbarr$, vanishes altogether,
making it possible for ice to spontaneously engulf the AFP.
We note that the precise supercooling at which engulfment occurs must depend 
not only on the engulfment barrier, but also the pre-exponential constant and the observation timescale~\cite{farag2023engulfment}; 
however, given the sensitive dependence of the engulfment barriers on $\dt$ (Figure~\ref{fig:Schematic}C),
here we approximate $\dts$ to be supercooling at which the barrier vanishes.
Indeed, using non-equilibrium simulations quenched to different temperatures, we find that ice growth is arrested for $\dt<\dts$, whereas AFP engulfment is observed for $\dt>\dts$ (Figure~S5).

%
To understand how AFP engulfment barriers vary with supercooling, 
we recognize that $\Gbarr$ is strongly correlated with $\gm$, 
the maximum value of the slope, $\glam \equiv dG/d\lambda$, 
which quantifies the the resistance to engulfment.
As $\dt$ is increased, both $\Gbarr$ and $\gm$ decrease until they vanish at $\dts$ (Figure~\ref{fig:Schematic}C).
The dependence of $\gm$ on $\dt$ is remarkably simple, 
and can be described well by the linear relationship, $\gm(\dt)=-\mg\dt+\cg$, so that $\dts = \cg/\mg$.
This key finding raises a number of interesting questions: 
(i) why is $\gm$ a linear function of $\dt$; 
(ii) is this trend expected for all separations, $L$, and for different AFPs; and if so,
(iii) how do $\mg$ and $\cg$, and correspondingly $\dts = \cg/\mg$, depend on AFP structure and separation?
%

\begin{figure*}[tbph]
\centering
\includegraphics[width=0.95\textwidth]{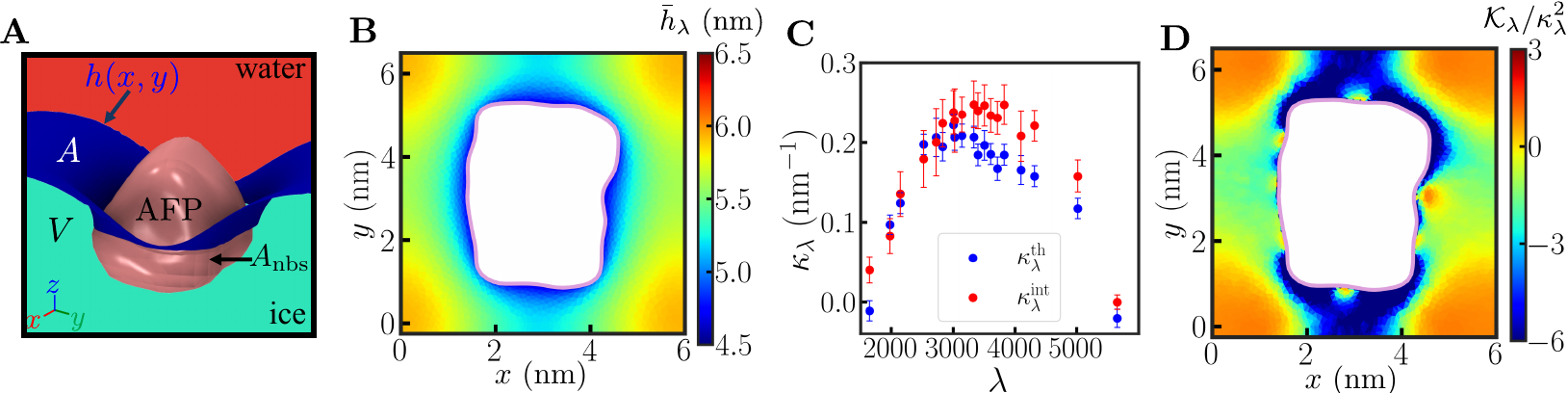}
\caption{
Shape of the ice-water interface during the engulfment of sbwAFP.
(A) Illustrating the interface, $h(x,y)$ (blue), separating the ice (cyan) and water (red) phases. 
The volume of the ice phase, $V$, the area of the ice-water interface, $A$, 
and the area of the NBS in contact with ice, $A_{\rm nbs}$, 
are highlighted along with the AFP NBS (pink).
(B) The simulated ice-water interface profile, $\hstar(x,y)$, is shown for a partially engulfed AFP with $\lambda \approx 3200$ ice-like waters, and the three-phase contact line is highlighted in pink.
(C) As the number of ice-like waters, $\lambda$, is increased from its bound state value, 
the magnitude of the mean curvature, $\kaplam$, increases from 0,
 reaches a maximum value of $\kapm = 0.22 \pm 0.02 $~nm$^{-1}$,
then decreases to 0 when the AFP is engulfed by ice.
The mean curvatures obtained from the simulated interfacial profiles, $\kaplam^{\rm int}$, 
agree well with the corresponding theoretical estimates, $\kaplam^{\rm th} = (\rho/2\gamma)(\glam+\dm)$. 
(D) The Gaussian curvature field, $\mathcal{K}_\lambda(x,y)$, normalized by the square of the mean curvature, shown for $\lambda \approx 3200$, highlights that the ice-water interface has a complex shape and displays regions of both positive and negative Gaussian curvature.
}
\label{fig:Engulfment}
\end{figure*}

\section*{Shape of the Ice-Water Interface}
To answer the questions raised above, 
we make use of macroscopic interfacial thermodynamics~\cite{gennes2004capillarity,remsing2015pathways,prakash2016spontaneous}, 
which provides the following expression for
the free energy, $\Gint$, of a system 
that has an ice-water interface profile, $h(x,y)$, and 
$\lambda$ ice-like waters at a supercooling, $\dt$:
\begin{equation}
\Gint([h], \lambda; \dt) = -\dm \rho V + \gamma A + \dgam A_{\rm nbs} + \lagmult (\lambda - \rho V),
\label{eq:macro}
\end{equation}
where $\dm \propto \dt$ is the chemical potential difference between liquid water and ice;
$\rho$ and $V$ are the density and volume of ice, respectively;
$\gamma$ and $A$ are the ice-water interface tension and area, respectively;
$\dgam$ is the difference between the NBS-ice and NBS-water surface tensions, 
and quantifies NBS ice-phobicity; 
$A_{\rm nbs}$ is the area of the NBS in contact with ice; and
$\lagmult$ is the Lagrange multiplier that constrains $\rho V$ to $\lambda$ (Figure~\ref{fig:Engulfment}A).
Because the values of $V$ and $A$ depend on the ice-water interface profile, 
the corresponding free energy, $\Gint$, is a functional of $h(x,y)$.
Equation~\ref{eq:macro} highlights that the engulfment free energetics stem from a complex interplay between the thermodynamic driving force, $\dm$, which favors ice growth, 
the ice-water interface tension, $\gamma$, which seeks to minimize the interfacial area, 
and the ice-phobicity of the NBS, $\dgam$, which disfavors engulfment.
%

To understand the implications of this interplay, we recognize that the 
mean-field 
ice-water interface,
which minimizes $\Gint([h],\lambda;\dt)$, 
must be a constant mean curvature surface, 
i.e., the sum of its principal curvatures must be the same for all $(x,y)$, 
and that 
the magnitude of the mean curvature, $\kaplam$, 
must be equal to $(\rho / 2\gamma)(\glam + \dm)$,  
as shown in the \SI.
Indeed, the average interface profile, $\hstar(x,y)$, obtained from our simulations, 
for a partially engulfed AFP with $\lambda \approx 3200$, 
is not flat, but has a negative mean curvature, as shown in Figure~\ref{fig:Engulfment}B.
The average interface profiles and mean curvature maps for a wide range of $\lambda$-values 
are included in the \SI~(Figure~S9),
along with the detailed procedure for obtaining them from our simulations.
In Figure~\ref{fig:Engulfment}C, we plot the magnitude of the interfacial curvature, $\kaplam$, 
as a function of $\lambda$, and find that it displays a maximum;
$\kaplam$ increases from zero at $\lamb \approx 1600$ (bound state at $\dt=0$), 
reaches its maximum, $\kapm = 0.22 \pm 0.02$~nm$^{-1}$ at $\lamm \approx 2900$, 
then decreases to eventually vanish at $\lambda \approx 5700$ (engulfed state). 
In Figure~\ref{fig:Engulfment}C, we also compare the curvature obtained from the simulated interface profiles, $\kaplam^{\rm int}$, to the corresponding theoretical prediction, $\kaplam^{\rm th} \equiv (\rho / 2\gamma)(\glam + \dm)$, and find good agreement between the two; 
here, $\glam$ was estimated using the simulated free energies, $G(\lambda; \dt)$ (Figure~\ref{fig:Schematic}B).

The close relationship between interfacial curvature and the resistance to engulfment, 
$\kaplam = (\rho / 2\gamma)(\glam + \dm)$,
implies that the maximum resistance to engulfment, $\gm$, ought to be determined by 
the maximum interfacial curvature, $\kapm$, that the AFP NBS can sustain through:
\begin{equation}
    \gm = -\dm + \bigg( \frac{2 \gamma}{\rho} \bigg) \kapm.
    \label{eq:gm_km}
\end{equation}
Importantly, Equation~\ref{eq:gm_km} sheds light on the linear 
relationship, $\gm(\dt) = -\mg \dt + \cg$, 
which was observed in Figure~\ref{fig:Schematic}C.
In particular, because $\dm \approx ( \Delta h / \Tm ) \dt$, where $\Delta h$ is the enthalpy of melting, 
and $(2\gamma/\rho) \kapm$ is largely independent of $\dt$, as shown in the \SI,
Equation~\ref{eq:gm_km} not only explains the 
linear dependence of $\gm$ on $\dt$, 
but also sheds light on the physical significance of the fit parameters, 
suggesting that $\mg = \Delta h / \Tm$ and $\cg = (2\gamma/\rho) \kapm$.
Indeed, the values of $\mg$ and $\cg$, obtained in Figure~\ref{fig:Schematic}C, 
are in good agreement with our estimates of $\Delta h / \Tm$ and $(2\gamma/\rho) \kapm$, as shown in the \SI.

By using macroscopic theory, we are thus able to answer the questions posed at the end of the previous section, and recognize that:
(i) the linear dependence of $\gm$ on $\dt$ stems from Equation~\ref{eq:gm_km} and $\dm \propto \dt$;
(ii) we expect this trend for all AFPs and separations; and
(iii) we expect $\mg \approx \Delta h/ \Tm$ to be a constant, and
$\cg$ to be influenced by AFP structure and separation through 
the maximum interfacial curvature, $\kapm$, that pinned AFPs are able to sustain.
Moreover, because $\gm$ vanishes at the critical supercooling, $\dts$, these insights into $\mg$ and $\cg$ result in the following expression for $\dts$:
\begin{equation}
    \dts = \frac{\cg}{\mg} = \frac{(2\gamma/\rho) \kapm}{\Delta h/ \Tm} = 2 \zeta \kapm,
\label{eq:k_dts}
\end{equation}
where $\zeta \equiv \gamma \Tm / \rho \Delta h$ is a physicochemical property of water,
and for the water model we use, $\zeta = 37.6$~K-nm~\cite{Abascal2005, espinosa2016ice}.
By plugging this value of $\zeta$ 
and 
$\kapm = 0.22 \pm 0.02$~nm$^{-1}$ (Figure~\ref{fig:Engulfment}C)
into Equation~\ref{eq:k_dts}, 
we obtain $\dts = 16.5 \pm 1.5$, which agrees well with our estimate of $\dts \approx 18$~K (Figure~\ref{fig:Schematic}C). 
We note that Equation~\ref{eq:k_dts} has the form of a Gibbs-Thomson equation,
i.e., it relates the supercooling, $\dts$, at which AFP engulfment occurs,
to the maximum interfacial curvature, $\kapm$, that pinned AFPs can sustain.
%

%
Given the importance of $\kapm$ in determining $\dts$, we now 
turn our attention to how the molecular characteristics of the NBS 
and the separation, $L$, between bound AFPs 
influence their ability to sustain highly curved ice-water interfaces.
Underpinned by the assumption that the ice-water interface has a regular shape 
(e.g., spherical cap) whose radius scales as $L$, 
it has been suggested that the interfacial curvature, $\kapm$, 
ought to vary inversely with $L$, i.e., $\kapm \propto L^{-1}$.
When combined with a Gibbs-Thompson equation (e.g., Equation~\ref{eq:k_dts}), 
this assumption suggests the widely believed scaling: $\dts \propto L^{-1}$~\cite{raymond1977adsorption, Knight:2001db, kristiansen2005mechanism, nada2012antifreeze_review_paper_MDsim, drori2015dts_L_dependance, bianco2020antifreeze, kuiper2015biological, midya2018operation, kozuch2018combined, naullage2018controls, gerhauser2021detailed}.
To test this assumption and interrogate whether the ice-water interfaces observed in our simulations (Figures~\ref{fig:Engulfment}B~and~S9) 
have regular shapes,
we plot the ratio of the Gaussian curvature field, $\mathcal{K}_\lambda(x,y)$, to the square of the mean curvature, $\kaplam^2$,
in Figures~\ref{fig:Engulfment}D~and~S9. 
If the interface were part of a sphere or a cylinder, the ratio ought to be 1 or 0, respectively, for all $(x,y)$.
In contrast, we find that the interfaces have widely varying Gaussian curvatures that take on 
both positive and negative values.
These findings highlight that AFP-pinned ice-water interfaces are not associated a unique radius of curvature, but have complex shapes, 
thereby calling into question the assumption that $\kappa \propto L^{-1}$, 
and correspondingly, the widely believed scaling, $\dts \propto L^{-1}$.

\section*{How Interfacial Curvature Scales with AFP Separation}
\begin{figure}[tbph]
\centering
\includegraphics[width=0.5\textwidth]{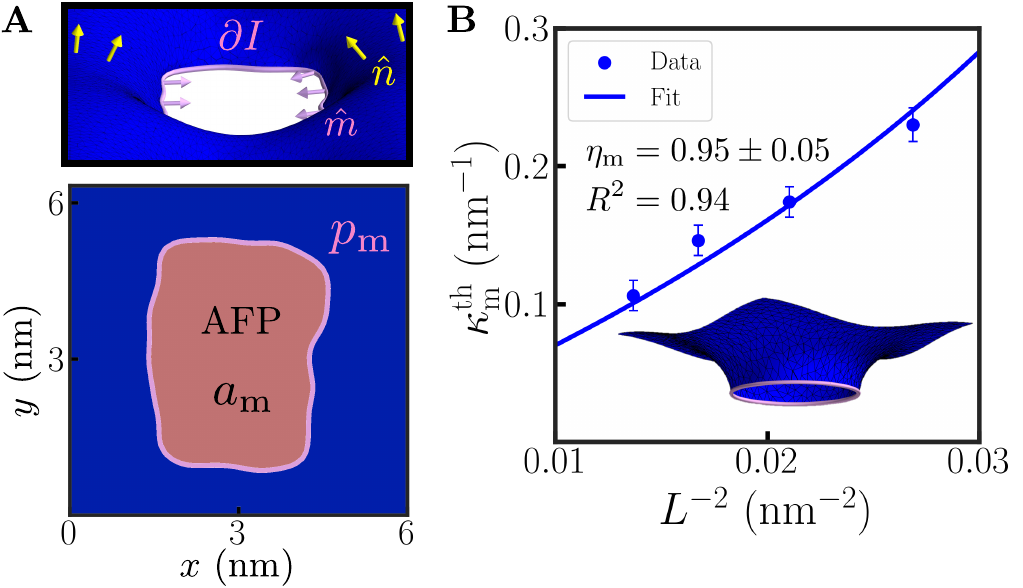}
\caption{
Variation of interfacial curvature with AFP separation. 
(A) The ice-water interface (blue) and the three-phase contact line, $\conl$ (pink), are illustrated in the top panel;
the unit vectors normal to the interface, $\nhat$ (yellow arrows), 
and to the contact line, $\mhat$ (purple arrows), are also shown for select points.
The $xy$ plane containing the contact line is shown in the bottom panel.
The perimeter of the contact line, $\Pm$, and the area enclosed by it, $\am$, are highlighted.
(B) The highest interfacial curvature that AFPs can sustain, $\kapm$, is shown as a function of $\Lsqinv$.
The simulation data (symbols) agree well with the best fit to Equation~\ref{eq:keta} (line).
Interestingly, the effectiveness, $\etam$, of the sbwAFP NBS at pinning the ice-water interface and resisting engulfment by ice is close to its optimal value, i.e., $\etam \approx 1$.
The inset illustrates that as $\etam \to 1$, the interface near $\conl$ becomes perpendicular to the far-field interface.
}
\label{fig:gradient}
\end{figure}

How does the maximum interfacial curvature, $\kapm$, sustained by bound AFPs, 
depend on their separation and on the molecular characteristics of the NBS?
To answer this question, we first recognize that the mean curvature 
at any point on the ice-water interface, $(x,y) \in \intf$,
is related to the divergence of $\nhat(x,y)$, the unit vector normal to the interface (Figure~\ref{fig:gradient}A), 
through $2 \kapm = \nabla \cdot \nhat$. 
Then, by using the divergence theorem, 
which equates the surface integral of $\nabla \cdot \nhat$ to 
the flux of $\nhat$ leaving the surface,
we get: $\iint_{\intf} \nabla \cdot \nhat~dxdy = \oint_{\conl} \nhat \cdot \mhat~ds$,
where $ds$ is the differential line element along the three phase contact line, $\conl$, 
and $\mhat$ is the unit vector normal to $\conl$ (Figure~\ref{fig:gradient}A).
Given that the ice-water interface is a constant mean curvature surface, 
i.e., $\kapm$ is independent of $(x,y)$, the surface integral 
simplifies to 
$2 \kapm (L_x L_y - \am)$, where $\am$ is the area enclosed by three phase contact line, $\conl$.  
Furthermore, because $\nhat$ and $\mhat$ are unit vectors, 
the line integral $\oint_{\conl} \nhat \cdot \mhat~ds$ is 
bounded by the perimeter of contact line, $\Pm \equiv \oint_{\conl}ds$, 
prompting us to define $\etam \equiv \oint_{\conl} \nhat \cdot \mhat~ds / \Pm$, 
such that $0 \le | \etam | \le 1$.
We note that $\etam$ depends solely on the molecular characteristics of the NBS;
$\mhat$ depends on its shape, 
whereas $\nhat$ depends on both its geometry and its chemistry (through $\Delta \gamma$).
Using $\etam$, the divergence theorem can then be written as: $2 \kapm (L_x L_y - \am) = \etam \Pm$.
From a physical standpoint, this equation represents a balance between the net upward force that the curved interface exerts on the contact line and the downward force that the ice-phobic NBS exerts on $\conl$.
Upon rearranging, we obtain: 
\begin{equation}
    \kapm = \frac{\etam \Pm}{2(\Lsq - \am)},
\label{eq:keta}
\end{equation}
%
where $L \equiv \sqrt{L_x L_y}$ is the geometric mean of $L_x$ and $L_y$.
Equation~\ref{eq:keta} highlights that $0 \le \etam \le 1$ quantifies the ability 
of the NBS to pin the contact line and sustain high interfacial curvatures.
Equation~\ref{eq:keta} also highlights that $\kapm$ is proportional to AFP size, as quantified by $\Pm$.
Importantly, Equation~\ref{eq:keta} predicts that for sufficiently large separations (i.e., $\Lsq \gg \am$), 
$\kapm$ scales as $\Lsqinv$.
%

To verify this scaling, 
we systematically vary the dimensions of our simulation box, 
and estimate the maximum curvature, $\kapm$, that can be sustained by the NBS,
as detailed in the \SI.
In Figure~\ref{fig:gradient}B, we plot $\kapm$ as a function of $\Lsqinv$, and find that 
the data (symbols) are in excellent agreement with the best fit to Equation~\ref{eq:keta} (line);
$\etam$ was used as the fit parameter, and
estimates of $\Pm$ and $\am$ were 
obtained from the shape of the AFP, as described in the \SI.
Our results thus refute the widely accepted scaling, $\kapm \propto \Linv$~\cite{raymond1977adsorption, Knight:2001db, kristiansen2005mechanism, nada2012antifreeze_review_paper_MDsim, drori2015dts_L_dependance, bianco2020antifreeze, kuiper2015biological, midya2018operation, kozuch2018combined, naullage2018controls, gerhauser2021detailed}, 
which is underpinned by the assumption of regular interface shapes, 
and suggest that interfacial curvature should instead obey the scaling, $\kapm \propto \Lsqinv$, 
in agreement with approaches that allow for complex interface shapes~\cite{sander2004kinetic,farag2023free}.
The best fit to Equation~\ref{eq:keta} also enables us to estimate $\etam$.
Interestingly, our estimate of $\etam$ for sbwAFP is close to its upper bound, 
i.e., $\etam \approx 1$, suggesting that the NBS of sbwAFP is particularly adept 
at pinning the contact line and sustaining highly curved ice-water interfaces.
We note that in the limit of $\etam \to 1$, the interface normal near the contact line 
becomes perpendicular to the direction of ice growth (Figure~\ref{fig:gradient}B, inset)~\cite{kristiansen2005mechanism}.
%

Collectively, these findings enable us to address how AFP structure and separation influence $\dts$.
In particular, by combining equations~\ref{eq:k_dts} with \ref{eq:keta}, we obtain:
\begin{equation}
\boxed{    
	\dts = \zeta \cdot \frac{ \etam \Pm }{ \Lsq - \am }
}.
\label{eq:dts}
\end{equation}
Equation~\ref{eq:dts} highlights that $\dts$ increases with AFP size (as quantified by $\Pm$) 
and that in contrast with the widely-held belief, 
$\dts$ scales as $L^{-2}$ (for $L^2 \gg \am$) and not $L^{-1}$~\cite{raymond1977adsorption, Knight:2001db, kristiansen2005mechanism, nada2012antifreeze_review_paper_MDsim, drori2015dts_L_dependance, bianco2020antifreeze, kuiper2015biological, midya2018operation, kozuch2018combined, naullage2018controls, gerhauser2021detailed}.
We also find that the NBS of sbwAFP is optimal at resisting engulfment by ice.
To what extent do these interesting insights translate to other AFPs?
How effective are other ice-binding proteins, obtained from different organisms, at resisting engulfment by ice? 
Finally, how sensitive is the $\etam$ of an AFP to mutations on its NBS?
%

\begin{figure*}[tbhp]
\centering
\includegraphics[width=0.95\textwidth]{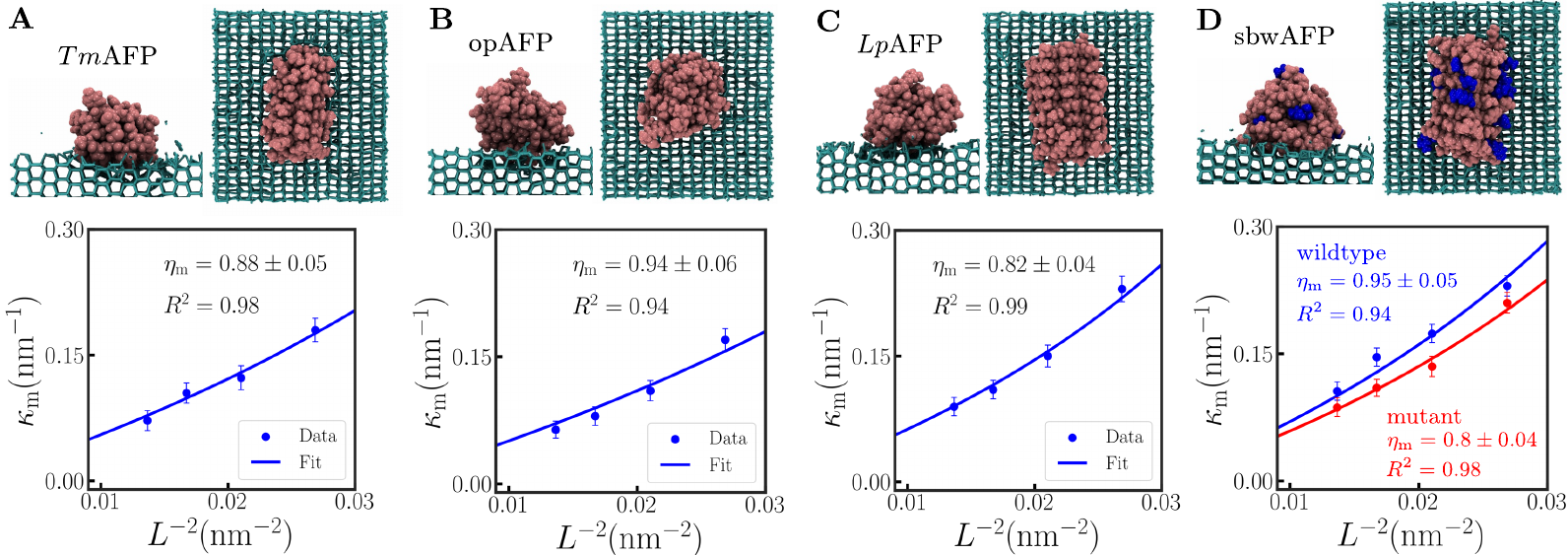}
\caption{
Characterizing the effectiveness of different AFPs at resisting engulfment by ice. 
Front and top views of the NBS of (A) an insect AFP ({\it Tm}AFP), (B) a fish AFP (opAFP), (C) a plant AFP ({\it Lp}AFP), and (D) wild-type sbwAFP are shown.
The AFPs are colored pink, whereas ice is shown in cyan (liquid waters are hidden for clarity); 
the charged residues on the sbwAFP NBS are shown in blue.
For each AFP, the variation of the maximum interfacial curvature, $\kapm$, with $\Lsqinv$ (symbols)
is found to be in excellent agreement with the corresponding fit to Equation~\ref{eq:keta} (lines); 
the fits also enable estimation of $\etam$, which quantifies the effectiveness of the AFPs at resisting engulfment.
Interestingly, the NBS of diverse naturally-occurring AFPs are nearly optimal at resisting engulfment, i.e., $\etam > 0.8$.
We also find that the $\etam$ of sbwAFP does not decrease substantially when all its charged residues are mutated to alanines, suggesting that the shape of the NBS play an important role in determining $\etam$. 
}
\label{fig:etam}
\end{figure*}

\section*{Diverse AFPs Excel at Resisting Engulfment} 
To interrogate the effectiveness of different naturally-occurring AFPs at resisting engulfment by ice, 
we now estimate $\etam$ for three more AFPs, which are expressed by a variety of organisms 
and span a wide range of activities: 
{\it Tm}AFP, a hyperactive insect AFP from the {\it Tenebrio molitor} beetle (Figure~\ref{fig:etam}A);
opAFP, a moderately active fish AFP from ocean pout (Figure~\ref{fig:etam}B); and 
{\it Lp}AFP, a hypoactive grass AFP from {\it Lolium perenne} (Figure~\ref{fig:etam}C).
For each of these diverse AFPs, we estimate the maximum interfacial curvature, $\kapm$, 
for four different AFP separations, and fit $\kapm$ vs $\Lsqinv$ to Equation~\ref{eq:keta} (Figures~\ref{fig:etam}A-C).
Interestingly, we find that $\etam > 0.8$ for all the AFPs, suggesting that these naturally-evolved proteins 
are proficient at resisting engulfment by ice.
A description of these simulations and the corresponding analysis is included in the \SI~(Figures~S11 -- S13).
%

The ability of an AFP to bind ice is remarkably sensitive to the chemical template presented by its IBS~\cite{mutagenesis1, mutagenesis3, mutagenesis4, mutagenesis7}. 
Is the effectiveness of an AFP at resisting engulfment by ice similarly sensitive to NBS chemistry or is the shape of the NBS more important in determining $\etam$?
To answer this question, we introduce mutations into the NBS of the wild-type sbwAFP.
In particular, because diverse AFPs have been observed to 
feature charged residues on their NBS~\cite{AFPs_from_diverse_organisms_paper_3_also_NBS_charge}, 
we interrogate their importance by mutating all 13 charged residues (8 positive, 5 negative) 
on the sbwAFP NBS into non-polar alanines (Figure~\ref{fig:etam}D).
Although these mutations drastically change the NBS chemistry, 
they do not appreciably alter the folded structure of the AFP, as shown in the \SI~(Figure~S14).
Interestingly, we find that the $\etam$ of the mutant sbwAFP ($\etam = 0.8 \pm 0.04$) 
is only somewhat lower than that of the wild-type ($\etam = 0.95 \pm 0.05$),
such that the mutant with its non-polar NBS remains proficient at pinning the ice-water interface (Figure~\ref{fig:etam}D).
Thus, although the charged residues on the sbwAFP NBS may have functional relevance (e.g., to enhance solubility), our findings suggest that they are not critical to resisting AFP engulfment by ice.
A detailed description of these calculations is included in the \SI~(Figure~S15).
%

The relative insensitivity of $\etam$ to mutations on the sbwAFP NBS is consistent with 
recent work by Marks et al.~\cite{marks2023characterizing}, who showed that 
both non-polar surfaces and polar surfaces, which are not lattice-matched with ice, can be ice-phobic.
Our finding that mutations on the NBS do not drastically alter $\etam$, and thereby $\dts$, 
also lends support to the practice of uncovering the AFP IBS 
by identifying mutants that significantly suppress $\dts$~\cite{mutagenesis1, mutagenesis3, mutagenesis4, mutagenesis7}.
Finally, we note that in introducing numerous charge-to-alanine mutations, we dramatically alter the chemistry of the sbwAFP NBS, but leave its shape largely unchanged.
Our findings thus suggest that the shape of the NBS may be more important in conferring engulfment resistance to the AFP than the presence of charged residues.
%

\section*{Conclusions and Outlook}
To inhibit the growth of ice crystals, antifreeze proteins (AFPs) must bind to ice, 
then resist engulfment by ice~\cite{raymond1977adsorption}.
The highest supercooling, $\dts$, for which AFPs are able to resist engulfment, 
thus determines the range of temperatures over which they are active~\cite{drori2014ice_dts_conc_annealing}. 
To understand how $\dts$ depends on the separation, $L$, 
between bound AFPs and on their molecular characteristics, 
here we characterize the free energetics of AFP engulfment using specialized molecular simulations. 
We find that at low supercooling, $\dt$, engulfment is impeded by large free energy barriers, 
which decrease as $\dt$ increases, and vanish above a critical supercooling, $\dts$, 
resulting in spontaneous engulfment.
By using interfacial thermodynamics, we then illustrate how 
engulfment barriers stem from a competition between 
the driving force for ice growth and 
the aversion of the AFP non-binding side (NBS) for ice,
which results in the ice-water interface being pinned to the NBS
and developing a (constant) mean curvature. 
We further show that the critical supercooling, $\dts$, is proportional to 
the highest interfacial curvature, $\kapm$, that bound AFPs can sustain, 
and that the ice-water interface adopts complex shapes 
with regions of both positive and negative Gaussian curvatures.
By using principles of differential geometry, 
we then relate $\kapm$ to the separation, $L$, between bound AFPs, 
and show that: $\dts \approx \zeta \etam \Pm \Lsqinv$,
where 
$\zeta$ is a physico-chemical property of water, 
$\Pm$ is the girth of the AFP, and
$0 \le \etam \le 1$ quantifies the effectiveness of the NBS at resisting engulfment by ice.
%

%
Interestingly, we find that $\etam$ is close to its optimal value, i.e., $\etam \approx 1$, 
for naturally-occurring AFPs obtained from a wide variety of organisms.
We also find that to be proficient at sustaining highly curved ice-water interfaces 
and resisting engulfment by ice,
an AFP must be able to pin the interface to be nearly perpendicular to the far-field interface~\cite{kristiansen2005mechanism}.
By mutating numerous charged residues to non-polar alanines, 
we further find that $\etam$ is relatively insensitive to 
the presence of charged residues on the sbwAFP NBS.
This finding suggests that the shape of the NBS 
may be more important than its chemistry in determining $\etam$.
We also find that $\dts$ is proportional to NBS size, $\Pm$, suggesting that protein engineering approaches, 
which do not alter the IBS, but enhance the size of the NBS, 
can nevertheless increase $\dts$ by providing enhanced resistance to engulfment.
Indeed, AFPs conjugated with bulky green-fluorescence protein tags have been observed to have higher thermal hysteresis (TH) activities than the corresponding non-tagged variants~\cite{deluca1998antifreeze, AFP_ice_binding, drori2014ice_dts_conc_annealing}.
%

In contrast with the widely-held belief that $\dts \sim \Linv$, 
here we show that $\dts \sim \Lsqinv$, i.e., $\dts$ is a more sensitive function of $L$.
By combining non-equilibrium fluorescence experiments with a reaction-diffusion model,
Thosar et al.~\cite{thosar2023accumulation} recently showed that the accumulation of AFPs at the ice-water interface~\cite{drori2014ice_dts_conc_annealing, kamat2021diffusion}, 
and thereby the average separation, $L$, between bound AFPs,
is not limited by the diffusion of AFPs,
but by the slow kinetics of AFP binding to ice~\cite{marshall2004a, meister2018antifreeze_JACS}.
Collectively, these findings suggest that the TH activity of an AFP ought to be 
sensitive to small changes in its IBS that influence its binding to ice and thereby $L$.
Indeed, by using protein engineering techniques to enlarge the IBS of {\it Gr}AFP,
Davies and co-workers were able to substantially increase its TH activity~\cite{scholl2023protein};
conversely, random mutations on the IBS often lead to a complete loss of TH activity~\cite{mutagenesis1,mutagenesis7}. 

Our findings also suggest that organizing bound AFPs into ordered arrays may result in greater TH activities.
In particular, because $\dts \sim \Lsqinv$, the first engulfment events must occur in regions with the highest separations;
such engulfment events further increase separation and are thus self-propagating~\cite{kristiansen2005mechanism, hansen2014edge, farag2023engulfment}.
Consequently, TH activity is not dictated by the average AFP separation, $L_{\rm avg}$, 
but by the highest AFP separation, $L_{\rm max}$,
i.e., TH $\sim L_{\rm max}^{-2} \sim \alpha^{-2} L_{\rm avg}^{-2} \sim \alpha^{-2} \theta$, 
where $\alpha \equiv L_{\rm max} / L_{\rm avg}$ and $\theta$ is the surface concentration of bound AFPs.
Thus, for a given $\theta$, the more heterogeneous the distribution of AFP separations (higher $\alpha$), 
the lower the TH activity (by a factor of $\alpha^2$).
Indeed, the $\dts$-values of hyperactive AFPs obtained using simulations, 
which feature a periodic array of bound AFPs (i.e., $\alpha \to 1$),
tend to be substantially larger than the corresponding 
experimentally-measured TH activities.
For example, we find that $\dts = 7.9$~K for {\it Tm}AFP molecules separated by $L = 7.7$~nm, 
whereas Drori et al. report a TH activity of $0.52$~K for {\it Tm}AFP molecules with an average AFP separation of $7.9$~nm~\cite{drori2015dts_L_dependance}, suggesting $\alpha \approx 3.8$.
Uncovering novel experimental protocols and/or AFP designs that organize bound AFPs (and lower $\alpha$) may thus offer a promising route to substantially greater TH activities.

%
Although the supercooling, $\dts$, at which an AFP is engulfed by ice 
is most closely related to the TH activity of hyperactive AFPs, such as sbwAFP or {\it Tm}AFP (which bind all ice planes),
it also informs the TH activity of moderately active AFPs, such as opAFP (which do not bind the basal plane)~\cite{AFP_ice_binding_Ran_paper_PNAS, drori2014ice_dts_conc_annealing}.
In particular, moderately active AFPs must not only bind to the prism (and/or pyramidal) planes, 
and arrest their growth, but also progressively shrink the exposed basal plane~\cite{knight2009ice};
thus, for moderately active AFPs, resisting engulfment is necessary, but not sufficient to arrest ice growth.  
Interestingly, most AFPs are also potent ice recrystallization inhibitors
and resisting engulfment is also a prerequisite for recrystallization inhibition~\cite{budke2014quantitative,bachtiger2021atomistic,tas2023nanoscopy}.
Thus, resisting engulfment by ice plays a central role in diverse AFP functions; 
by elucidating how $\dts$ depends on the separation between bound AFPs 
and on their molecular characteristics (Equation~\ref{eq:dts}), 
we hope that our findings will help uncover the corresponding structure-function relationships.
We also hope that the conceptual and methodological framework introduced here will find use in understanding crystal growth inhibition in other systems, such as gas hydrates~\cite{Storr:2004fd}, 
and in designing AFP-like inhibitors for a wide variety of crystals~\cite{rimer2010crystal, shtukenberg2017crystal}. 
%

\begin{acknowledgement}
This material is based upon work supported by the U.S. Department of Energy (DOE), Office of Science, Office of Basic Energy Sciences, under Award Number DE-SC0021241. 
S.M.M. was supported by a DOE Computational Science Graduate Fellowship (DE-FG02-97ER25308).
AJP gratefully acknowledges early-career awards from the Alfred P. Sloan Research Foundation (FG-2017-9406) and the Camille \& Henry Dreyfus Foundation (TC-19-033), as well as 
numerous insightful conversations with Valeria Molinero and Ran Drori.

\end{acknowledgement}

\bibliography{ms}

\end{document}


\newpage

\section{Molecular Simulations of Antifreeze Proteins (AFPs)}

\subsection{Simulation parameters}
All molecular dynamics simulations were performed using GROMACS version 2016.3~\cite{Berendsen1995-ec}.
The equations of motion were integrated using the leap-frog algorithm with a time step of 2~fs, 
and periodic boundary conditions were used in all dimensions. 
To simulate systems at a particular temperature, the stochastic velocity-rescale thermostat~\cite{Bussi2007-rp} was employed with a time constant of 0.5~ps, whereas system pressure was maintained using the anisotropic Parrinello-Rahman barostat~\cite{Nose1983-gd, Parrinello1981-hv} with a time constant of 4~ps; 
fluctuations in the simulation box dimensions were allowed only in the $z$-direction.
Water molecules were represented explicitly using the TIP4P/Ice model~\cite{Abascal2005-tm} and 
the AFPs were modeled using the Amber99SB force field~\cite{Hornak2006-ux}.
The Lorentz-Berthelot mixing rules were used to obtain the cross-parameters for Lennard-Jones interactions.
Both the Lennard-Jones and the short-range electrostatic interactions were truncated at 1~nm, 
and long-range electrostatic interactions were treated using the particle mesh Ewald (PME) method~\cite{Darden1993-nt}.
Water hydrogens were constrained using the SETTLE algorithm~\cite{Miyamoto1992-ru}, whereas bonds involving protein hydrogens were constrained using the LINCS algorithm~\cite{Hess1997-vj,Hess2008-gp}. 
To restrain the position and orientation of the AFP, 
the positions of all protein heavy atoms within 0.5~nm of its center of mass
were restrained using a relatively soft spring constant of 1000~kJ/mol/nm$^2$.
By restraining only a small fraction of heavy atoms (roughly 20~atoms or 1~\% of the total) near the protein's center of mass, we are able to localize the AFP without restraining any of its surface atoms, 
i.e., all protein atoms on the IBS and the NBS can fluctuate freely.

\begin{figure}[H]
\vspace{-0.1in}
\centering
\hspace{-2cm}
\includegraphics[width=11cm]{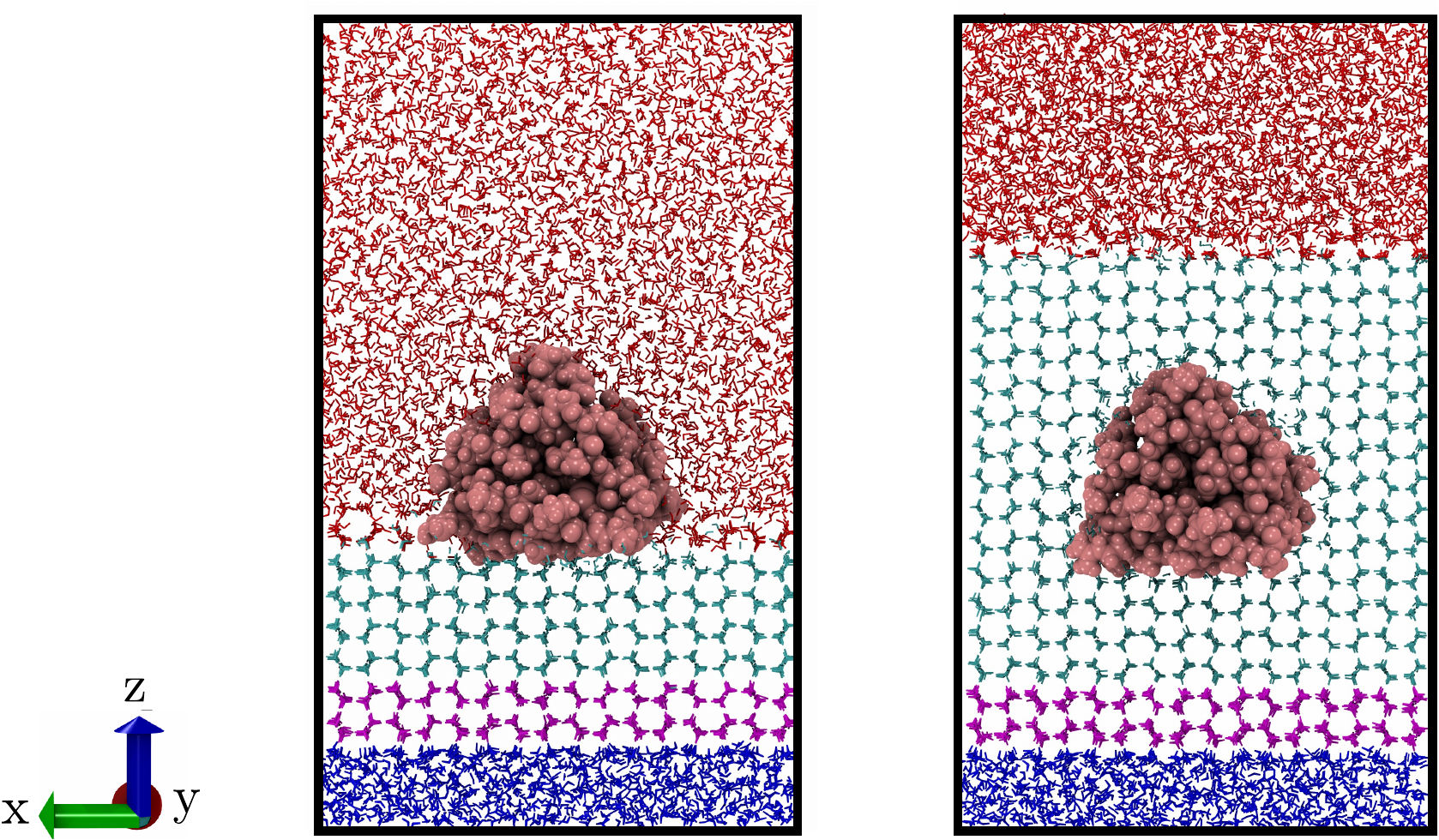}
\caption{
(A) Simulation snapshot of sbwAFP (pink) bound to the primary prism plane of ice.
The water oxygens shown in blue and purple are position restrained, 
resulting in liquid-like and ice-like boundary slabs, respectively, 
which ensure that ice grows in the $+z$-direction.
The unrestrained (mobile) ice and liquid water molecules are shown in cyan and red, respectively.
(B) Simulation snapshot of sbwAFP engulfed by ice, which serves as the initial configuration for biased simulations. 
The snapshots were visualized using the Visual Molecular Dynamics (VMD) package~\cite{HUMP96}.
}
\vspace{-0.1in}
\label{fig:set-up}
\end{figure}

\subsection{Simulation set-up}
We studied the following four AFPs: (i) sbwAFP from the spruce budworm ({\it Choristoneura Fumiferana}, Isoform 501; PDB ID: 1M8N), (ii) {\it Tm}AFP from {\it Tenebrio molitor} (PDB ID: 1EZG), (iii) opAFP from ocean pout ({\it Zoarces americanus}, PDB ID: 1HG7), and (iv) {\it Lp}AFP from {\it Lolium perenne} (PDB ID: 3ULT)~\cite{leinala2002-vc, Liou2000-tt, Antson2001-he, Middleton2012-wj}. 
The protein systems were prepared for simulation using the `pdb2gmx' utility of the GROMACS molecular dynamics package~\cite{Berendsen1995-ec}. 
The majority of our calculations were performed using sbwAFP with the simulation box dimensions, $L_x$ and $L_y$, 
chosen to be roughly 6~nm (Figure 1A) to ensure that the protein heavy-atoms were at least 1~nm away from the box edges.
Simulation boxes with $L_x$ and $L_y$ ranging from 6~nm to 9~nm were used to characterize the variation of interfacial curvature with AFP separation (Figures~3B and~4).
To set up our simulations, the GenIce package~\cite{Matsumoto2018-cu} was first used to generate a 2~nm thick slab of ice (Figure~\ref{fig:set-up}A, cyan) with its primary prism ($10\bar{1}0$) plane normal to the $z$-direction. 
The Packmol package~\cite{Martinez2009-bh} was then used to judiciously place the AFP of interest in the simulation box.
In particular, the AFP was centered in the $x$ and $y$ directions, and along the $z$-direction, the AFP was placed less than 0.5~nm from the ice slab with its IBS facing the primary prism plane;
the azimuthal orientation of the AFP was chosen to ensure that the chemical groups on the IBS (e.g., threonine hydroxyls) were aligned with the hydroxyls of the ice slab.
The protein was then solvated in a roughly 7~nm thick slab of water (Figure~\ref{fig:set-up}A, red) using the `solvate' utility of GROMACS; sodium or chloride ions were added, if necessary, to maintain the charge-neutrality of the system. 
Lastly, two 1~nm thick position-restrained slabs of liquid water (Figure~\ref{fig:set-up}A, blue) and ice (Figure~\ref{fig:set-up}A, purple) were introduced at the bottom of the simulation box to break translational symmetry and ensure that ice grows in the +$z$ direction.
To prevent intrusion of the mobile water molecules (Figure~\ref{fig:set-up}A, cyan and red) 
into these boundary slabs, the spring constants for the restraining potentials were chosen to be 
40,000~kJ/mol/nm$^2$ and 4,000~kJ/mol/nm$^2$ for the liquid water and ice slabs, respectively.
The solvated AFP was then position-restrained, as described above, 
and energy-minimized using the steepest-descent algorithm, 
followed by a 0.5~ns long NVT simulation (at $T=298$~K), 
and a 2~ns long NPT simulation (at $T=298$~K and $P = 1$~bar).
To facilitate binding of the AFP to the ice-water interface, 
the position restraints on the AFP were relaxed, 
and a 20~ns long simulation was run at $T=298$~K and $P=1$~bar 
in the presence of a biasing potential that prevents the ice slab from melting, as described in sec.~2B below.
The final configuration is shown in Figure~\ref{fig:set-up}A.
%

\section{Characterizing the Free Energetics of AFP Engulfment}

To prepare the initial configurations for our simulations as well as 
to subsequently characterize the free energetics of AFP engulfment, 
we bias the number of ice-like mobile waters in our simulations.

\subsection{Biasing the approximate number of ice-like waters.}
\begin{figure}[htb]
\vspace{-0.15in}
\centering
\includegraphics[width=\textwidth]{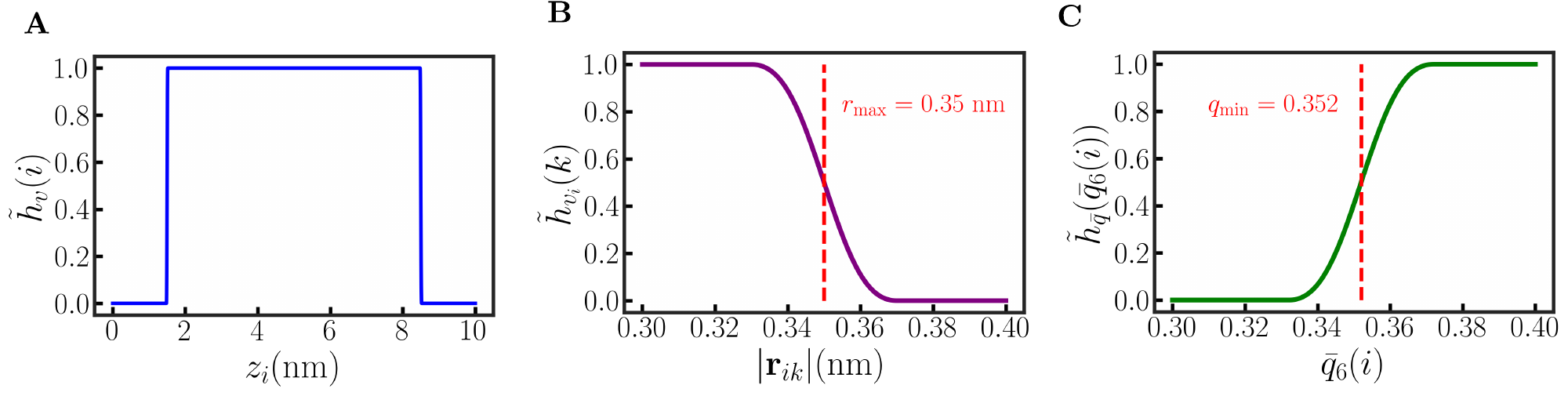}
\caption{
Functional forms of the indicator functions used to calculate the number of ice-like molecules, $M$, in an observation volume, $v$, of interest. 
(A) The indicator function $\tilde{h}_{v}(i)$ determines whether or not water molecule $i$ is inside $v$.
(B) The indicator function $\tilde{h}_{v_{i}}(k)$ determines whether water $k$ is a nearest neighbor of water $i$.
(C) The indicator function $\tilde{h}_{\bar{q}}(\bar{q}_6(i))$ determines whether the local orientational order of water $i$ resembles that of ice.
}
\label{fig:indicator_func}
\vspace{-0.in}
\end{figure}

To bias the number of ice-like waters using molecular dynamics simulations, 
the order parameter being biased must not only be capable of discriminating between liquid water and ice, 
but also be a continuous and differentiable function of particle positions~\cite{patel2011quantifying}.
To this end, we define the order parameter, $M$, corresponding to the approximate number of ice-like molecules within a observation volume $v$ in our system as follows~\cite{marks_characterizing_2023}:
\begin{equation}
M \equiv \sum_{i=1}^{N_{\rm w}} \tilde{h}_{v}(i) \cdot \tilde{h}_{\bar{q}}(\bar{q}_6(i)),
\label{eq:M}
\end{equation}
where the sum is performed over the $N_{\rm w}$ mobile waters in the simulation box; 
the coarse-grained function $\tilde{h}_{v}(i)$ indicates whether a water molecule $i$ is in an observation volume of interest, $v$;  
the indicator function, $\tilde{h}_{\bar{q}}(\bar{q}_{6}(i))$, is used to classify whether a water molecule, $i$, is ice-like, and its argument, $\bar{q}_{6}(i)$ is closely related to the local order parameter introduced by Lechner and Dellago~\cite{Lechner2008-rl}. 
In particular, $\bar{q}_{6}(i)$ is defined as:
\begin{equation}
\bar{q}_{6}(i) \equiv \sqrt{ \frac{4\pi}{13} \sum_{m=-6}^{6} | \bar{q}_{6m}(i) |^2} \text{,}
\label{eq:q6bar}
\end{equation}
where
\begin{equation}
\bar{q}_{6m}(i) \equiv \frac{1}{1 + \tilde{N}_{\mathrm{nn}}(i)} \cdot \Bigg[ q_{6m}(i) + \sum_{k=1}^{N_{w}} \tilde{h}_{v_{i}}(k) q_{6m}(k) \Bigg].
\label{eq:q6mbar}
\end{equation}
Here, $\tilde{h}_{v_{i}}(k)$ indicates whether water $k$ is a nearest neighbor of water $i$, 
$\tilde{N}_{\mathrm{nn}}(i) \equiv \sum_{k \ne i}^{N_{\rm w}} \tilde{h}_{v_{i}}(k)$ is the coarse-grained number of nearest neighbors of water $i$, and
\begin{equation}
    q_{6m}(i) \equiv \frac{1}{\tilde{N}_{{\rm{nn}}}(i)}\sum_{k = 1}^{N_{\rm w}} \tilde{h}_{v_{i}}(k) Y_{6m}(\mathbf{r}_{ik}),
\label{eq:q6m}
\end{equation}
where $Y_{6m}(\mathbf{r}_{ik})$ is the spherical harmonic function, and its argument, $\mathbf{r}_{ik}$, is the vector from the oxygen of water $i$ to that of water $k$. 
To ensure that the indicator functions, $\tilde{h}_{v}(i)$, $\tilde{h}_{v_{i}}(k)$, and $\bar{h}_{\bar{q}}$ are continuous and differentiable functions of particle positions, they are obtained by integrating Gaussian functions of width $\sigma$ that are truncated and shifted at $2\sigma$ following refs.~\cite{patel2010fluctuations,marks_characterizing_2023}. 
The indicator functions are plotted in Figure~\ref{fig:indicator_func}. 
The observation volume, $v$, is chosen to be box-spanning in the $x$ and $y$ directions, 
and spans from $z_{\rm min} = 1.5$~nm to $z_{\rm max} = 8.5$~nm along the $z$ direction, 
so that $\tilde{h}_{v}(i)$ increases continuously from 0 to 1 
(over a distance of $z_{\rm min/max} \pm 2\sigma$ with $\sigma=0.01$~nm) 
as water molecule $i$ approaches the boundaries of $v$, as shown in Figure~\ref{fig:indicator_func}A; 
$z_i$ denotes the $z$-co-ordinate of water $i$.
Figure~\ref{fig:indicator_func}B shows the indicator function $\tilde{h}_{v_{i}}(k)$ used to determine the nearest neighbors of water $i$. 
As the distance, $r_{ik}$, between water $i$ and its putative neighbor $k$ increases above $r_{\rm max} = 0.35$~nm, the value of $\tilde{h}_{v_{i}}(k)$ decreases smoothly from 1 to 0 (over $r_{\rm max} \pm 2\sigma$ with $\sigma=0.01$~nm).
Figure~\ref{fig:indicator_func}C shows the indicator function $\tilde{h}_{\bar{q}}$, 
which determines whether the local orientational order of water $i$ resembles that in bulk ice. 
As $\bar{q}_{6}(i)$ crosses the value of $q_{\rm min} = 0.352$, 
which occurs with equal probability in bulk water and ice at 273~K,
the value of $\tilde{h}_{\bar{q}}$ increases smoothly from 0 to 1 (over $q_{\rm min} \pm 2\sigma$ with $\sigma=0.01$). 
The approximate number of ice-like waters, $M$, was biased, 
and the corresponding forces were distributed across the water molecules in the system 
using an in-house plug-in that was developed to interface with the PLUMED package~\cite{Bonomi2009-gf, Tribello2014-vh, PLUMED_consortium2019-jm}.
%

\subsection{Biased Simulations}
We performed biased simulations to prepare the initial configurations for subsequent calculations as well as to characterize the free energetics of AFP engulfment.
In particular, biased simulations were used to obtain initial configurations with the AFP bound to ice (Figure~\ref{fig:set-up}A) and the AFP engulfed by ice (Figure~\ref{fig:set-up}B). 
To obtain the former, a linear biasing potential, $\phi M$, was applied (with $\phi = -0.548$~kJ/mol) to prevent the melting (or growth) of the ice slab (Figure~\ref{fig:set-up}A, cyan).
To engulf the AFP, we initialized the system in the bound state configuration (Figure~\ref{fig:set-up}A), 
and ran a 20~ns long biased simulation using a linear potential, $\phi M$ (with $\phi = -1.1$~kJ/mol), which resulted in the engulfed configuration shown in Figure~\ref{fig:set-up}B.
Subsequent biased simulations for characterizing the free energetics of engulfment were initialized using the engulfed configuration (Figure~\ref{fig:set-up}B), 
and employed harmonic potentials, $\frac{\kappa}{2} (M-M^{*})^{2}$, 
with $M^{*}$ being varied systematically to sample the pertinent range of $M$-values. 
To mitigate the formation of defects, $M^{*}$ is decreased linearly to its desired value over 3~ns. 
All biased simulation are run for 15~ns, which includes 3~ns for varying $M^{*}$ to its final value, the subsequent 5~ns for equilibration, and the final 7~ns for production.
%

\subsection{Estimating the true number of ice molecules.}
\begin{figure}[htpb]
\vspace{-0.1in}
\centering
\includegraphics[width=0.8\textwidth]{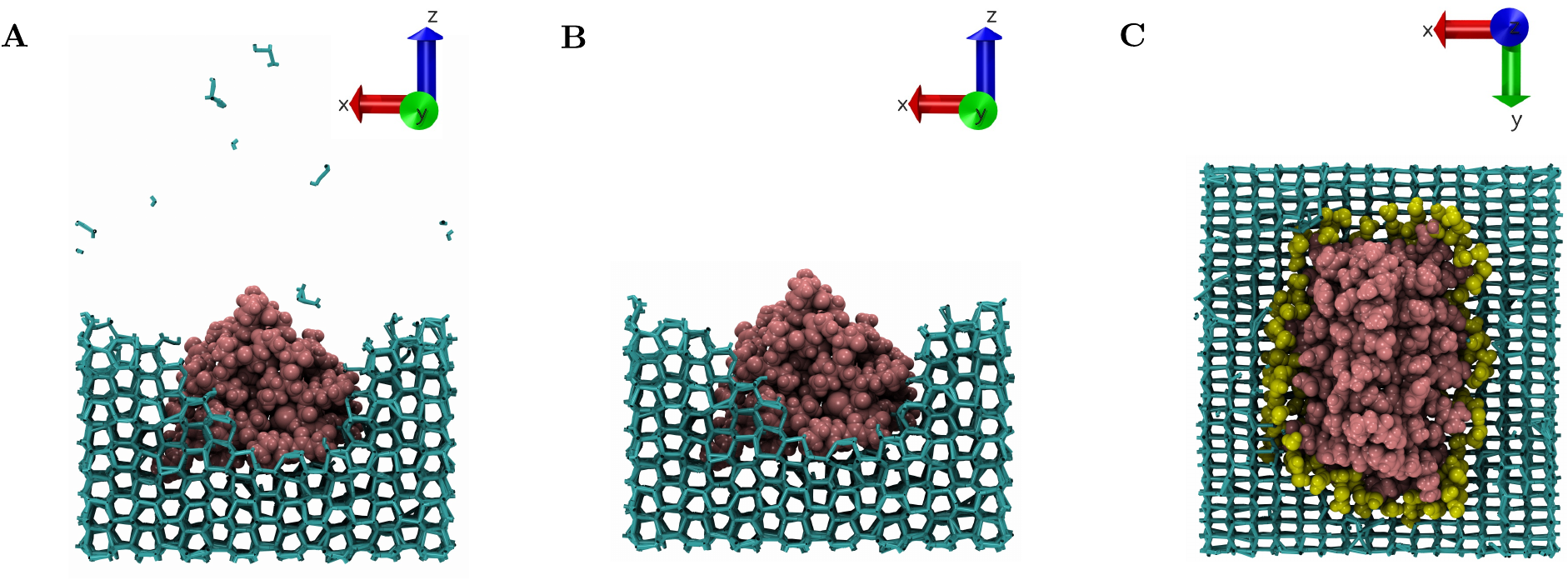}
\vspace{-0.1in}
\caption{
Estimating the true number of ice molecules, $\lambda$.
(A) Ice-like waters identified using the CHILL+ algorithm~\cite{Nguyen2015-fd} are shown (cyan) 
for a partially engulfed AFP (pink, front view), whereas liquid waters are hidden for clarity. 
A few molecules in bulk water are misidentified as ice-like.
(B) Ice molecules determined using Moller-Trumbore algorithm~\cite{Moller1997-de} to be above the instantaneous ice-water interface~\cite{Willard2010-pk} are filtered out. 
(C) Waters near the AFP (top view) that were reclassified to being ice-like are shown here in yellow.
} 
\label{fig:identify_lambda}
\vspace{-0.1in}
\end{figure}

Although $\bar{q}_{6}(i)$ and similar order parameters are excellent at discriminating between liquid water and ice, a small fraction of water molecules in the liquid phase will nevertheless have high $\bar{q}_6(i)$-values, and be incorrectly classified as ice-like, giving rise to false positives.
Moreover, classifying interfacial water molecules, i.e., those near solid surfaces, as ice-like and liquid-like is more challenging than classifying waters in the bulk because the interfacial waters are neighbor-deficient, 
i.e., they have fewer than the four nearest neighbors found in the bulk phases.
Thus, the order parameter, $M$, defined in Equation~\ref{eq:M} above, 
is unable to identify ice-like molecules in the vicinity of the AFP, resulting in false negatives.
To more accurately estimate the true number of ice molecules, $\lambda$, we first used the CHILL+ algorithm~\cite{Nguyen2015-fd} to identify the ice-like waters (hexagonal, cubic and interfacial ice) in our system and classified the remainder as liquid-like waters.
As shown in Figure~\ref{fig:identify_lambda}A, although CHILL+ is excellent at identifying molecules that belong to ice, it nevertheless misidentifies a few molecules that are clearly in bulk water.
To filter out the misidentified molecules,
we located the instantaneous ice-water interface following ref.~\cite{Willard2010-pk},
and water molecules that were determined to be above the interface
using the Moller-Trumbore algorithm~\cite{Moller1997-de} were reclassified as being liquid-like.
As shown in Figure~\ref{fig:identify_lambda}B, this procedure was able to successfully filter out the misidentified ice-like molecules seen in Figure~\ref{fig:identify_lambda}A.
Finally, to identify ice-like interfacial waters, i.e., those near the AFP, we used the following procedure: if an AFP hydration water (i.e., water within 0.6~nm of a protein heavy atom), 
which has been classified as liquid-like, is found to be within 0.55~nm of 8 or more ice-like waters, 
it is reclassified as being an ice-like water; 
this procedure is then repeated recursively until no further reclassification occurs.
As shown in Figure~\ref{fig:identify_lambda}C, this protocol enables identification of the ice-like waters near the AFP (yellow).

%

\subsection{Reweighting} 
\begin{figure}[htb]
\vspace{-0.1in}
\centering
\includegraphics[width=0.8\textwidth]{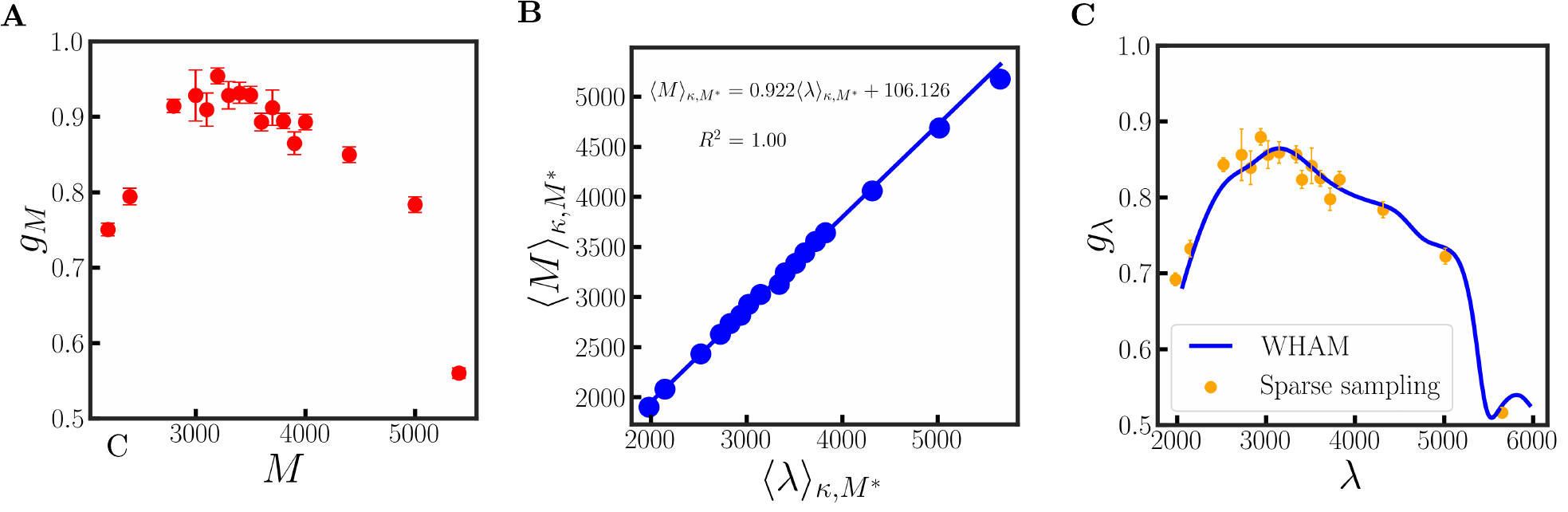}
\caption{
Estimating $g_{\lambda}$ for sbwAFP with $L \approx 6$~nm and $T = 298$~K. 
(A) At select values of $M = \langle M \rangle_{\kappa,M^*}$, $g_M$ is shown as a function of $M$.
(B) We find $\langle M \rangle_{\kappa,M^*}$ to be linearly correlated with $\langle \lambda \rangle_{\kappa, M^*}$ with the slope of the best linear fit being 0.92. 
(C) Estimates of $\glam$ obtained using sparse sampling (symbols) are in good agreement with those obtained using WHAM (line).
}
\vspace{-0.1in}
\label{fig:g}
\end{figure}

Although the approximate number of ice-like waters, $M$, is biased using a harmonic potential, $\frac{\kappa}{2} (M - M^*)^2$ in our simulations,
the correlation between $M$ and the true number of ice molecules, $\lambda$, 
enables us to indirectly sample a wide range of $\lambda$-values across our biased simulations~\cite{patel2011quantifying}.
We then reweight the biased simulations to obtain the free energetics of engulfment, $G(\lambda;\dt)$, and the corresponding resistance to engulfment, $\glam \equiv dG/d\lambda$.
For biased distributions with overlap in their sampled $M$ (and $\lambda$) values,
$G(\lambda;\dt)$ and $\glam$ were obtained using standard weighted histogram reweighting methods~\cite{Shirts2008-iv,Zhu2012-sv,patel2011quantifying}.
To estimate $\glam$ more efficiently for different separations, $L$, and AFP types, we additionally made use of the sparse sampling approach~\cite{Xi2018-po}, which does not require overlap between adjacent biased distributions.
In particular, we first obtained $g_M \equiv dG/dM$ at select values of $M=\langle M \rangle_{\kappa,M^*}$ using the relationship $g_M \big|_{M = \langle M \rangle_{\kappa,M^*}} = \kappa ( M^* -  \langle M \rangle_{\kappa,M^*} )$~\cite{Xi2018-po}, 
as shown in Figure~\ref{fig:g}A.
We then averaged $\lambda$ over our biased simulations to obtain estimates of $\langle \lambda \rangle_{\kappa,M^*}$, and found a strong linear correlation between $\langle M \rangle_{\kappa,M^*}$ and $\langle \lambda \rangle_{\kappa,M^*}$, as shown in Figure~\ref{fig:g}B.
Accordingly, we posited that $M$ and $\lambda$ were linearly related to one another with the proportionality constant, $M_{\lambda}$, being well-approximated by the slope of the linear fit to $\langle M \rangle_{\kappa,M^*}$ vs $\langle \lambda \rangle_{\kappa,M^*}$ (Figure~\ref{fig:g}B).
Finally, we estimated $\glam$ for $\lambda$-values corresponding to $\lambda = \langle \lambda \rangle_{\kappa,M^*}$ as:
\begin{equation}
g_{\lambda}\big|_{\lambda=\langle \lambda \rangle_{\kappa,M^*} } = g_{M} \big|_{M = \langle M \rangle_{\kappa,M^*} }  \cdot M_{\lambda}.
\label{eq:reweight}
\end{equation}
In Figure~\ref{fig:g}D, we compare our estimates of $\glam$ obtained using sparse sampling (Equation~\ref{eq:reweight}) against those obtained using WHAM~\cite{Shirts2008-iv,Zhu2012-sv}, 
and find excellent agreement between them.
The errors in our sparse sampling estimates of $\glam$ were obtained using error propagation 
with contributions from the standard errors in $\langle M \rangle_{\kappa,M^*}$ as well as errors in $M_\lambda$ (obtained using block bootstrapping). 
%

\section{Non-equilibrium Simulations}
\begin{figure}[htb]
\centering
\includegraphics[width=0.8\textwidth]{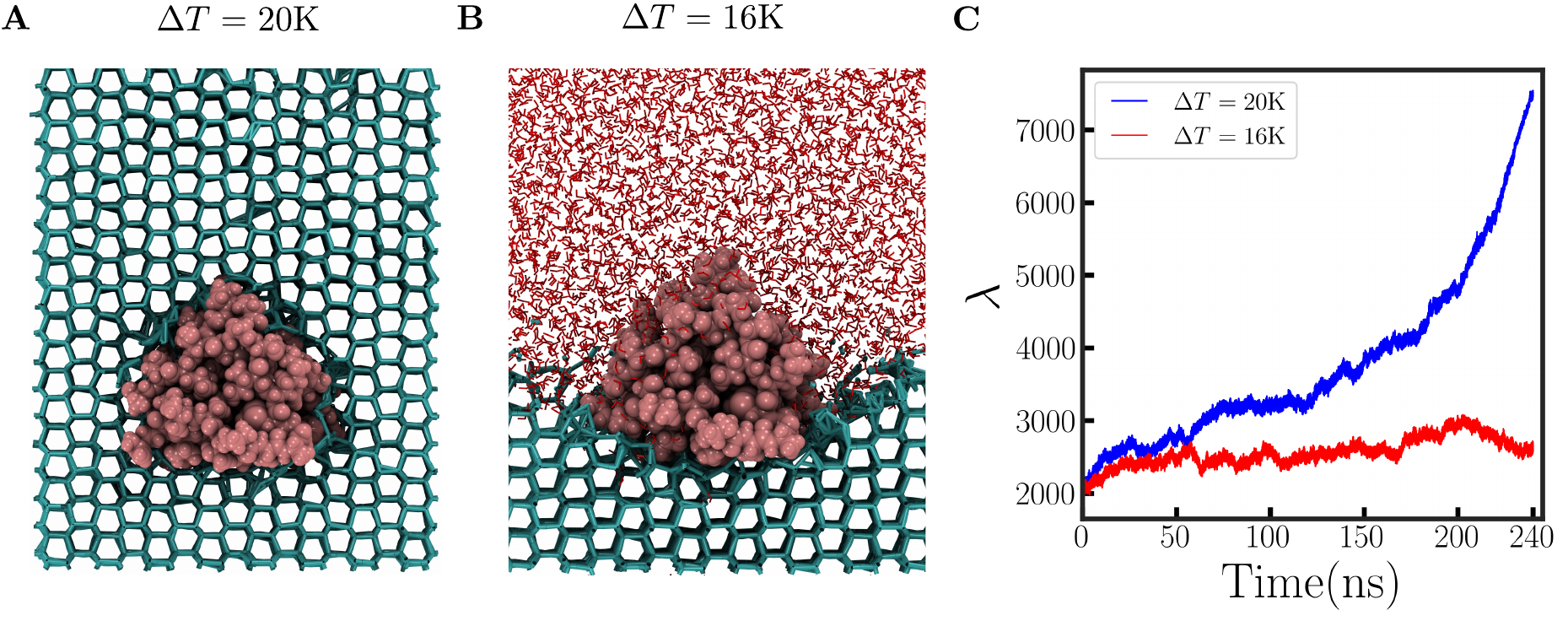}
\caption{
(A) When sbwAFP bound to the ice-water interface is quenched to a supercooling of $\dt = 20$~K, it is engulfed by ice, but (B) when it is quenched to a supercooling of $\dt = 16$~K, it arrests ice growth.
(C) The number of ice molecules, $\lambda$, is shown as a function of time 
for non-equilibrium simulations with a supercooling of $\dt = 20$~K (blue) and $\dt = 16$~K (red).
}
\label{fig:non-equi}
\end{figure}

To interrogate whether the supercooling at which sbwAFP (with $L \approx 6$~nm) is engulfed spontaneously by ice corresponds to the supercooling, $\dts = 18$~K, at which engulfment barriers in $G(\lambda;\dt)$ vanish, we performed unbiased non-equilibrium simulations wherein an sbwAFP molecule 
bound to the ice-water interface is quenched to different temperatures.
The system is initialized with the AFP bound to ice, as shown in Figure~\ref{fig:set-up}A. 
Then, at time $t=0$, the system is supercooled to either $\dt = 20$~K or $\dt = 16$~K,
and the corresponding non-equilibrium simulations are run for 240~ns; 
the final configurations of these simulations are shown in Figures~\ref{fig:non-equi}A and~\ref{fig:non-equi}B, respectively, and the time-dependence of the number of ice molecules during these simulations is shown in Figure~\ref{fig:non-equi}C.
These results highlight that sbwAFP arrests ice growth for $\dt < \dts$, but is engulfed by ice for $\dt > \dts$.
%

\section{Macroscopic Interfacial Thermodynamics}
%
According to macroscopic interfacial thermodynamics, the free energetics of a system, which has an ice-water interface profile, $h(x,y)$, contains $\lambda$ ice-like waters, and is at supercooling, $\dt$, is given by: 
\begin{equation}
G_{\rm th}([h],\lambda;\Delta T) = -\dm \rho V[h] + \gamma A [h] + \Delta \gamma A_{\rm nbs} + \mathcal{L}_\lambda (\lambda - \rho V[h]),
\label{eq:macro}
\end{equation}
where $V[h]=\iint h \,dx \,dy$ is the volume of ice in the system and $A[h]=\iint \sqrt{1+ h_x^2 + h_y^2} \,dx \,dy$ is the ice-water interfacial area.
As discussed in the main text, 
$\dm$ is the chemical potential difference between liquid water and ice,
$\rho$ is the density of ice, 
$\gamma$ is the ice-water interfacial tension, 
$\dgam$ quantifies NBS ice-phobicity, 
$A_\mathrm{nbs}$ is the surface area of the NBS in contact with ice; and
$\lagmult$ is a Lagrange multiplier that constrains $\rho V$ to $\lambda$.
Because we are interested in studying AFP engulfment, we assume that the AFP is bound to ice for all interfacial profiles, $h(x,y)$, and $\lambda$-values under consideration.  above.
Moreover, ice-phobicity can vary across different regions of the NBS, 
so $\Delta \gamma$ represents an average over the area of the NBS that is in contact with ice. 
Finally, because $A_{\rm nbs}$ does not depend on the shape of the interfacial profile, but only on the location of its boundaries, the functional derivative of $G_{\rm th}$ with respect to $h(x,y)$ is:
\begin{equation}
\frac{\delta G}{\delta h} = -\rho ( \mathcal{L}_\lambda + \Delta \mu ) \frac{\delta V}{\delta h} + \gamma \frac{\delta A}{\delta h} 
\label{eq:rearrange}
\end{equation}
Therefore, the mean-field interface, $\bar{h}(x,y)$, must obey:
\begin{equation}
\frac{\delta G}{\delta h} \bigg|_{\bar{h}} = 0 \implies \frac{\delta A/\delta h}{\delta V/ \delta h} \bigg|_{\bar{h}}= \frac{\rho (\mathcal{L}_\lambda + \Delta \mu)}{\gamma}.    
\label{eq:var_derivtive}
\end{equation}
By taking the functional derivatives of $V[h]$ and $A[h]$, we get $\delta V / \delta h = 1$ and:
\begin{equation}
    \frac{\delta A}{\delta h} = -\frac{[1+h_x^2] h_{yy} - 2 h_x h_y h_{xy} + [1+h_y^2] h_{xx}}{[1+ h_x^2 + h_y^2]^{3/2}} = 2 \kaplam(x,y),
\label{eq:curvature}
\end{equation}
where $\kaplam(x,y)$ is the negative of the mean interfacial curvature.
Combining Equations~\ref{eq:var_derivtive} and \ref{eq:curvature}, we obtain:
\begin{equation}
    \kappa_{\lambda}(x,y)\bigg|_{\bar{h}} = \frac{\rho}{2\gamma} \cdot (\mathcal{L}_\lambda + \Delta \mu).
\label{eq:kappa}
\end{equation}
Because the right hand side of this equation is independent of $x$ and $y$, the mean-field ice-water interface must be a constant mean curvature surface.

To illustrate the physical significance of the Lagrange multiplier, $\mathcal{L}_\lambda$, we recognize that the mean-field free energy of the system is given by $\bar{G}(\lambda; \dt) = G_{\rm th}([\bar{h}],\lambda;\dt)$, and the corresponding resistance to engulfment, $\glam \equiv d\bar{G}(\lambda; \dt)/d\lambda$ is:
\begin{equation}
    \glam = \frac{d}{d\lambda}G_{\rm th}([\bar{h}],\lambda;\dt) = 
    \int \int \cancelto{0}{ \frac{\delta G_{\rm th}}{\delta h}\bigg|_{\bar{h}} } \cdot \frac{d\bar{h}}{d\lambda} \cdot dx dy + \frac{\partial}{\partial \lambda}G_{\rm th}([\bar{h}],\lambda;\dt) = \mathcal{L}_\lambda + \frac{\partial \mathcal{L}_\lambda}{\partial \lambda} \cdot \cancelto{0}{( \lambda - \rho V[\bar{h}] )} = \mathcal{L}_\lambda.
\label{eq:lm}
\end{equation}
%
Finally, combining Equations~\ref{eq:kappa} and \ref{eq:lm}, we obtain an expression for the magnitude of the interfacial curvature:
\begin{equation}
    \kaplam = \frac{\rho}{2\gamma} \cdot (\glam + \Delta \mu).
\label{eq:kaplam}
\end{equation}
\section{Characterizing the shape of the ice-water interface}
In this section, we describe the methods used to characterize the shape of the ice-water interface, observed in our biased simulations, and to obtain the corresponding mean and normalized Gaussian curvature fields.
We illustrate the techniques using partially engulfed sbwAFP molecules that are separated by roughly 8~nm.

\subsection{Locating the interface of ice with water and/or the AFP}
\begin{figure}[htb]
\centering
\includegraphics[width=0.6\textwidth]{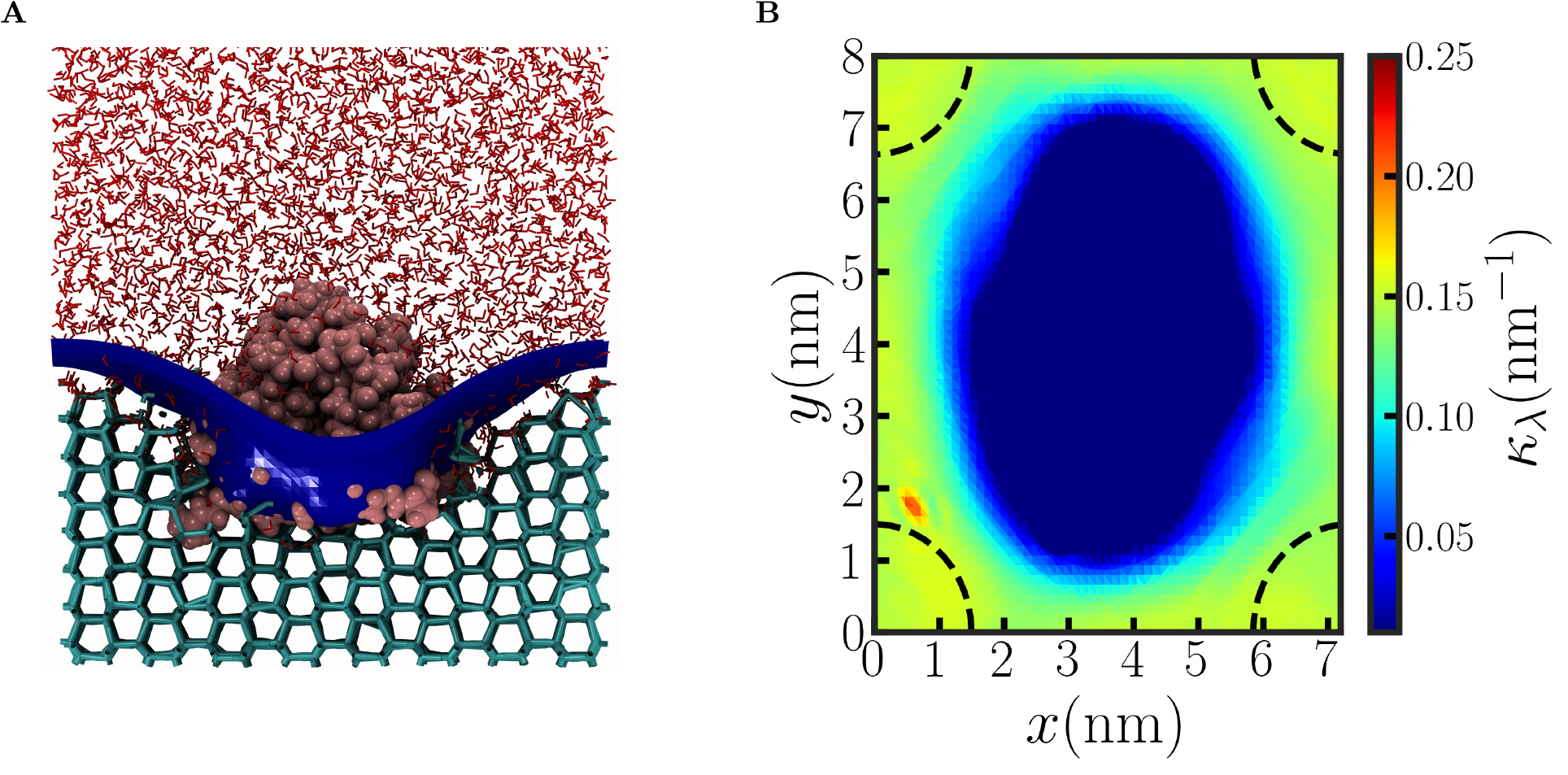}
\caption{
(A) The (average) interface (blue) separating ice (cyan) from liquid water (red) and the AFP (pink) is superimposed with a simulation snapshot. 
(B) The mean curvature map of the smoothed interface.
}
\label{fig:curvature_map}
\end{figure}

For every configuration in our biased simulations, we first identified the ice-like molecules using the procedure described in section 2C. 
Following ref.~\cite{Willard2010-pk}, we then obtained the coarse-grained density of ice-like molecules 
at every position, $\mathbf{r}$, in our simulation box position as: 
\begin{equation*}
    \rho(\mathbf{r}) = \sum_{i} \phi(| \mathbf{r} - \mathbf{r_{i}} | ),
\end{equation*}
where $\mathbf{r_{i}}$ corresponds to the position of the $i^{\rm th}$ ice-like water and the coarse graining function $\phi(r)$ is defined as: 
\begin{equation*}
    \phi(r) = (2\pi \xi^{2})^{-3/2} \exp(-r^{2}/2\xi^{2}),
\end{equation*}
with the coarse-graining length, $\xi = 0.24$~nm. 
The instantaneous ice density field, $\rho(\mathbf{r})$, was then averaged over the simulation trajectory, 
to obtain the averaged ice density field, $\bar{\rho}(\mathbf{r})$,
and the Marching cubes algorithm was used to obtain the interface of ice with liquid water and/or the AFP.
In particular, the interface was defined as the $\bar{\rho}(\mathbf{r}) = s$ iso-surface, with $s$ chosen to be 15~nm$^{-3}$ (i.e., roughly half the density of bulk ice). 
As shown in Figure~\ref{fig:curvature_map}A, the interface (blue) separates ice (cyan) from liquid water (red) and the AFP (pink).
%

\subsection{Smoothing the interface and estimating interfacial curvature}
Because curvature is sensitive to rough interfacial features, 
we first smoothed the interface using the implicit fairing algorithm, 
which retains the geometric characteristics of the interface (e.g., curvature)~\cite{Desbrun1999-uf}. 
In particular, 10 iterations of the implicit fairing algorithm were executed with a step size of 0.1~nm$^2$. 
We then used polynomial fitting to compute the mean curvature at every point on the smoothed interface~\cite{Cazals2005-ot}.
The mean curvature map thus obtained is shown in Figure~\ref{fig:curvature_map}B; 
its central region corresponds to the AFP-ice interface, 
whereas its peripheral regions correspond to the ice-water interface.
To obtain an estimate of the (constant) mean curvature associated with the ice-water interface, $\kaplam^{\rm int}$,
we averaged the interfacial curvature over the regions farthest from the AFP (dashed lines shown in Figure \ref{fig:curvature_map}B).

\subsection{Identifying the three-phase contact line}
\begin{figure}[htbp]
\centering
\includegraphics[width=1.\textwidth]{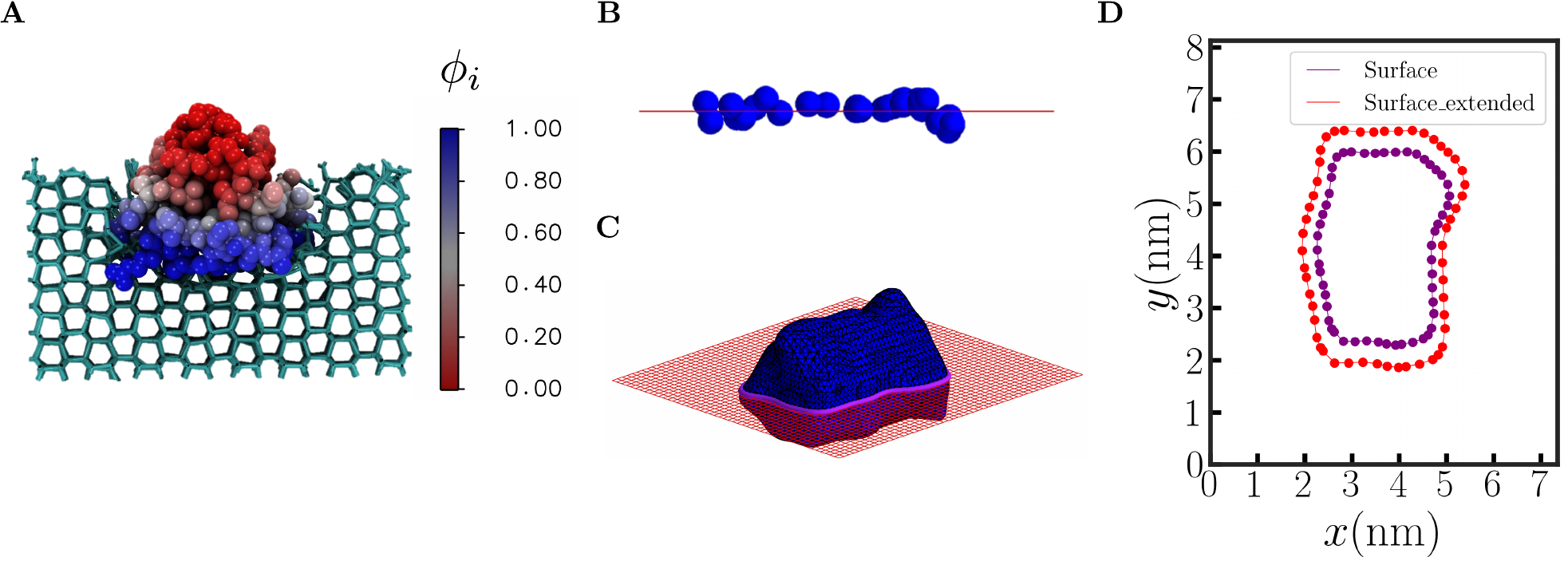}
\caption{
(A) Simulation snapshot denoting ice molecules (cyan) and AFP heavy atoms colored according to their $\phi_i$-values; the atoms colored in white ($0.5 \leq \phi_i \leq 0.6$) approximate the location of the three-phase contact line where the ice-water interface is pinned to the AFP.
(B) The positions of the atoms that fall on the three-phase contact line (blue) are fitted to a plane (red). 
(C) The intersection of the contact plane (red) with the surface mesh of protein (blue). 
(D) The intersection of contact plane with protein surface mesh (purple) is extended outwards by the size of a water molecule to obtain the three-phase contact line (red).
}
\label{fig:conl}
\end{figure}

Although the above procedure provides a faithful representation of the ice-water interface far from the AFP, the shape of the ice-water interface near the AFP is obfuscated by the proximity of the AFP-ice interface.
To address this challenge, we first identify the three-phase contact line, $\conl$.
To this end, 
we averaged the number of ice molecules within 1~nm of every AFP surface heavy atom, 
and normalized it by the corresponding quantity for the fully engulfed AFP.
A simulation snapshot of the partially engulfed sbwAFP is shown in Figure~\ref{fig:conl}A 
with every protein heavy atom $i$ colored according to the normalized number of ice molecules, $\phi_i$, in its vicinity; atom $i$ was deemed to be a contact line atom if $0.5 \leq \phi_i \leq 0.6$.
The $z$-coordinates of the contact line atoms were tightly clustered, so the atoms were fitted to a contact plane that is normal to the $z$-direction, as shown in Figure~\ref{fig:conl}B. 
The intersection between the contact plane and the protein surface (Figure~\ref{fig:conl}C) 
then provided an initial approximation of the contact line (Figure~\ref{fig:conl}D, purple). 
To obtain the protein surface mesh, an iso-surface of the average coarse-grained density field of all mobile waters, i.e., those belonging to both liquid water and ice, was used, as described in section~5A.
We note that an iso-surface of the coarse-grained density field of protein heavy atoms can also be used to obtain the protein mesh, and it results in a nearly identical estimate of the contact line.
Because the contact line resides on the ice-water interface (and not on the protein), 
we extended our initial estimate of the contact line outward by 0.32~nm (corresponding to the Lennard Jones $\sigma$ of the water oxygen) to obtain our final estimate of the contact line, $\conl$ (Figure~\ref{fig:conl}D, red). 

\begin{figure}[htbp]
\centering
\includegraphics[width=0.8\textwidth]{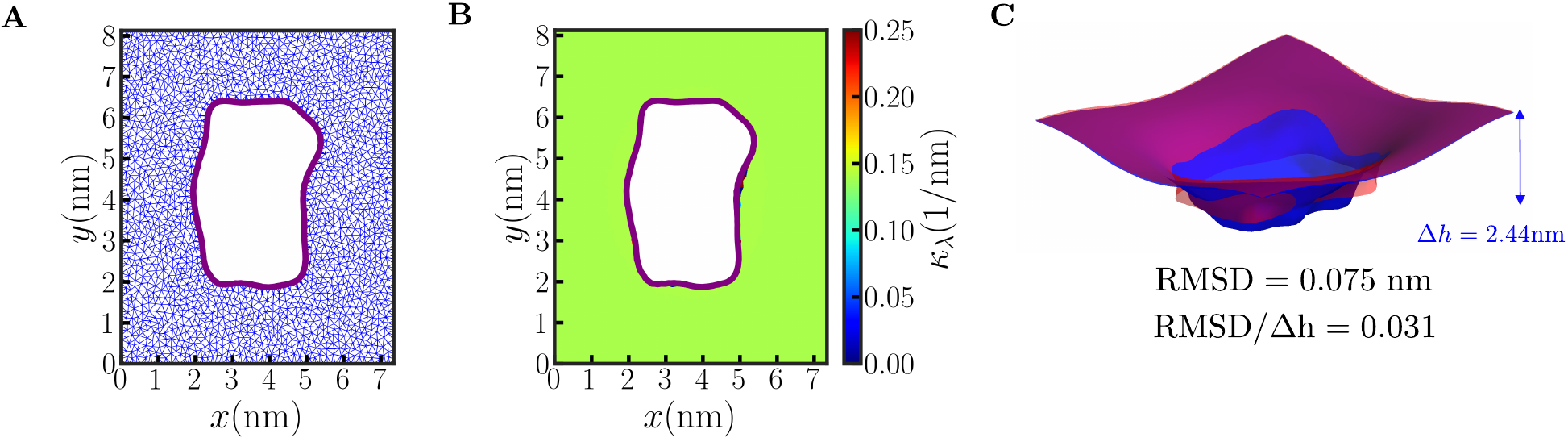}
\caption{
Identifying the ice-water interface near the contact line.
(A) A periodic flat mesh excluding the region enclosed by the contact line is generated using constrained Delaunay triangulation~\cite{cgal:eb-23a}. 
(B) The mesh is pinned at the contact line and evolved into a constant mean curvature surface using the Lagrange process~\cite{Lobaton2007-yr}; 
the curvature map of the converged mesh confirms that it is indeed a constant mean curvature surface. 
(C) The evolved surface (red) and the smoothed interface (blue, Figure~\ref{fig:curvature_map}A) agree well with one another. 
The root mean square difference (RMSD) between the surfaces is less than 0.1~nm, which is roughly 3\% of the range of interface heights, $\Delta h$. 
}
\label{fig:grow_surf}
\end{figure}
\subsection{Determining the ice-water interface near the contact line}
To determine the shape of the ice-water interface near the contact line, $\conl$, 
we used constrained Delaunay triangulation~\cite{cgal:eb-23a} 
to first construct a flat 2D mesh, which excludes the region enclosed by the contact line, 
and is periodic in the $x$ and $y$ directions (Figure~\ref{fig:grow_surf}A).
We then employed a Lagrange process~\cite{Lobaton2007-yr} to evolve the flat mesh into a constant mean curvature surface using $\kaplam^{\rm int}$ determined above (sec.~5B).
In particular, keeping the mesh pinned at the contact line, the positions of the rest of the mesh vertices were evolved using:
\begin{equation}
\frac{d\bar{r}_{i}}{dt} = -(\kappa_{i} - \kaplam^{\rm int}) ~ \hat{n}_{i},
\label{eq:lagrange_evolution}
\end{equation}
where $\bar{r}_{i}$ is the position of vertex $i$, $\kappa_{i}$ is the magnitude of the mean curvature at that position, and $\hat{n}_{i}$ is the corresponding unit vector normal to the interface.
Upon convergence (i.e., $d\bar{r}_{i} / dt = 0$), the generated interface satisfies $\kappa_{i}= \kaplam^{\rm int}$ for every mesh vertex $i$. 
The converged mean curvature field, thus obtained, is shown in Figure~\ref{fig:grow_surf}B.
As shown in Figure~\ref{fig:grow_surf}C, the corresponding ice-water interface (red) agrees well with the smoothed interface (blue, Figure~\ref{fig:curvature_map}A). 
We also computed the root mean square difference between the two surfaces (RMSD), 
and found it to be less than 0.1~nm or roughly 3~\% of the range of heights spanned by the ice-water interface, $\Delta h$ (Figure~\ref{fig:grow_surf}C).
%

\begin{figure}[htbp]
\centering
\includegraphics[width=0.7\textwidth]{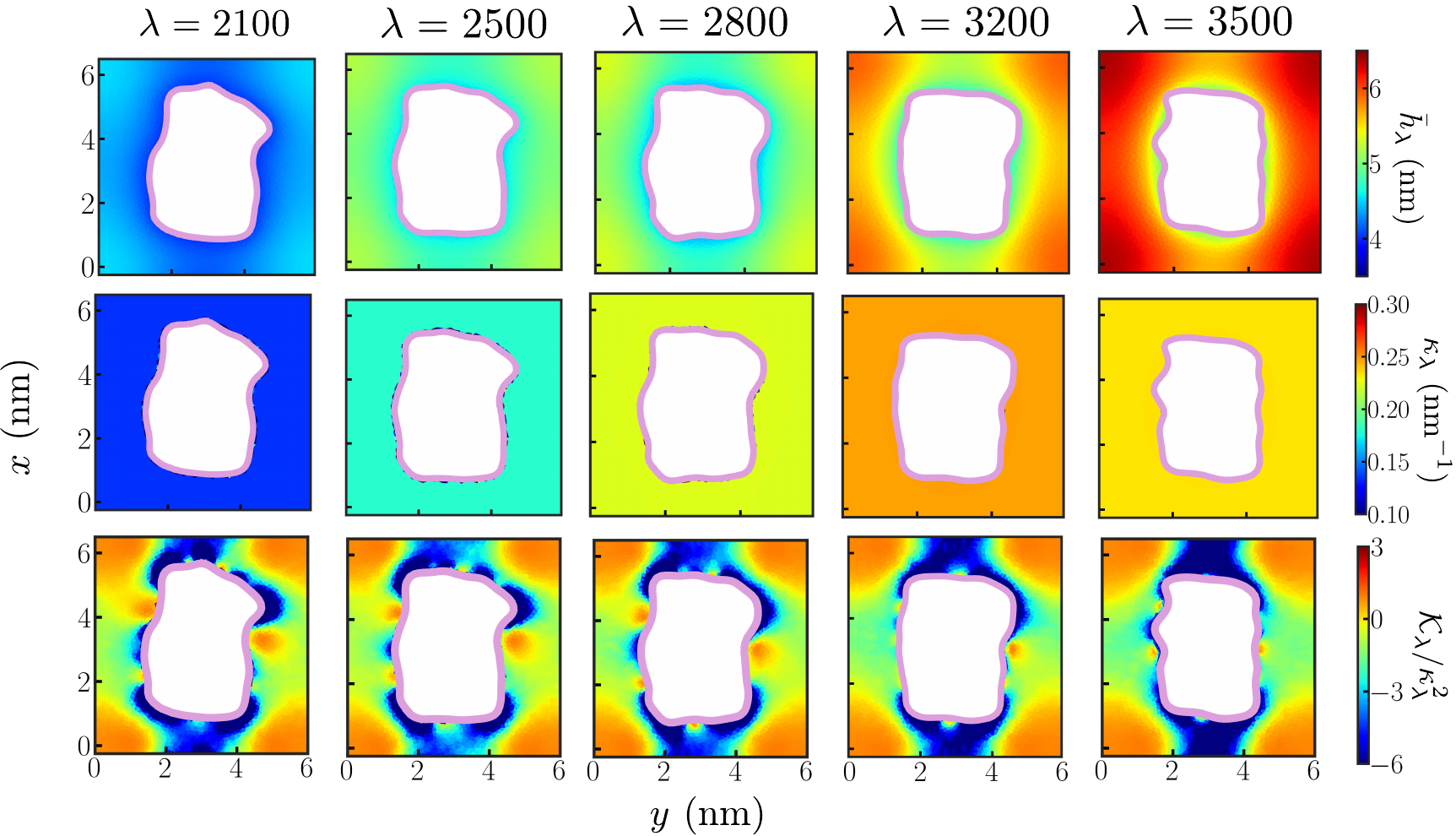}
\caption{
For sbwAFP molecules separated by roughly 6~nm, as $\lambda$ is increased, the evolution of the average ice-water interface profiles, $\hstar(x,y)$, is shown in the top row, the corresponding mean curvature fields, $\kaplam(x,y)$, are shown in the middle row, and the corresponding Gaussian curvatures, $\mathcal{K}_\lambda$, normalized by the square of the mean curvature, are shown in the bottom row.
In each case, $\kaplam(x,y)$ is independent of $(x,y)$ (middle row), highlighting that the ice-water interfaces are constant mean curvature surfaces.
However, the interfaces have complex shapes and display regions of both positive and negative Gaussian curvature (bottom row).
}
\label{fig:CMC}
\end{figure}

Using the above procedure to characterize the shape of the ice-water interface, both near and far from the AFP, we were able to obtain the interface profiles and curvature fields for a wide range of $\lambda$-values as sbwAFP molecules, separated by roughly 6~nm, undergo engulfment; see Figure~\ref{fig:CMC}.
%

\section{Variation of $\gm$ with $\dt$}
According to Equation~2 of the main text: $\gm = -\dm + (2 \gamma/\rho) \cdot \kapm$.
Upon taking the derivative of this equation with respect to the supercooling, $\dt$, we obtain: 
\begin{equation}
	\frac{d \gm}{d \dt} = - \frac{d \dm}{d \dt} + \frac{d}{d \dt} \bigg( \frac{2\gamma}{\rho} \bigg) \kapm \approx \frac{ \Delta h }{ T_{\rm m} } - 2 \kapm \frac{d}{dT} \bigg( \frac{\gamma}{\rho} \bigg), 
\label{eq:gm-Tdep}
\end{equation}
where we use the fact that $\dm \approx -(\Delta h/T_{\rm m}) \dt$ and that $d\dt = - dT$.
Because $\Delta h = 6.01$~kJ/mol and $T_{\rm m} = 273.15$~K, the first term on the right hand side is: $\Delta h/T_{\rm m} = 0.022$~kJ/mol-K.
Similarly, for our water model, i.e., TIP4P/Ice~\cite{Abascal2005-tm}, $\Delta h = 5.6$~kJ/mol and $T_{\rm m} = 271$~K, so that $\Delta h/T_{\rm m} = 0.02$~kJ/mol-K.
To estimate the second term on the right hand side, we note that the maximum curvature obtained for sbwAFP separated by $L = 6$~nm was roughly $\kapm = 0.22$~nm$^{-1}$ (Figure~2C), and that $\kapm$ only decreases for larger $L$ (varying as $\Lsqinv$).
Moreover, $d(\gamma/\rho)/dT = -0.0049$~nm-kJ/mol-K~\cite{Ickes2015-an} 
(and -0.00533~nm-kJ/mol-K for our water model~\cite{Espinosa2016-ko}),
so that the second term must be smaller than 0.0022~kJ/mol-K (for $L \ge 6$~nm).
Thus, the second term on the right hand side of Equation~\ref{eq:gm-Tdep} is roughly an order of magnitude smaller than the first, suggesting that $(2\gamma/\rho) \kapm$ is a weak function of $T$ relative to $\dm$,
and that the variation of $\gm$ with $\dt$ stems primarily from the (linear) variation of $\dm$ with $\dt$.

These arguments provide a basis for understanding the linear relationship, $\gm(\dt) = -\mg \dt + \cg$, observed in Figure~1C, and suggest that the fit parameters must be given by: $\mg = (\Delta h / T_{\rm m})$ and $\cg = (2\gamma/\rho) \kapm$.
We find that $\mg = 0.02~{\rm kJ/mol-K}$ (Figure~1C) is indeed in excellent agreement with $(\Delta h/T_{\rm m}) = 0.02~{\rm kJ/mol-K}$ of the TIP4P/Ice water model~\cite{Abascal2005-tm}.
Similarly, $\cg = 0.36$~kJ/mol (Figure~1C) agrees well with $(2\gamma/\rho)\kapm = 0.324$~kJ/mol, where $\gamma = 3.7 \times 10^{-5}$~kJ/m$^2$ and $\rho = 5 \times 10^4$~mol/m$^3$ for TIP4P/Ice~\cite{Abascal2005-tm,Espinosa2016-ko}, and $\kapm = 0.22$~nm$^{-1}$ (Figure~2C).
%
%

\section{Estimating $\etam$ for sbwAFP}
In Equation~4 of the main text, we showed that the maximum interfacial curvature, $\kapm$, that can be sustained by the AFP is given by: 
\begin{equation}
    \kapm = \frac{\etam \Pm}{2(\Lsq - \am)} = \frac{1}{2} \etam \Pm \frac{\Lsqinv}{(1 - \am \Lsqinv)},
\label{eq:si-keta}
\end{equation}
%
where 
$\etam$ is the  effectiveness of the NBS at resisting engulfment by ice, 
$\am$ and $\Pm$ are the area enclosed by the three phase contact line and its perimeter, respectively,
and $L$ is the geometric mean of the simulation box dimensions, $L_x$ and $L_y$.
To estimate $\etam$ for the AFP of interest, first we systematically varied the AFP separation, $L$ 
(by varying the dimensions of our simulation box), and uncovered how it influences the maximum curvature, $\kapm$, that can be sustained by the NBS.
Then, we obtained estimates of $\Pm$ and $\am$ using the contact line (Figure~\ref{fig:conl}D, red).
Finally, we fit the $\kapm$ vs $L$ relationship to Equation~\ref{eq:si-keta}, 
and obtained $\etam$ as the parameter, which provided the best fit to the data.
%

\subsection{Characterizing the variation of $\kapm$ with $L$}
To characterize how $\kapm$ varies with $L$, we performed biased simulations of AFP engulfment using different simulation boxes sizes.
To determine $\kaplam$ for a given simulation box size, $L$, 
we used Equation~\ref{eq:kaplam}, i.e., $\kaplam^{\rm th} = \frac{\rho}{2\gamma} \cdot (\glam + \dm)$.
In particular, by leveraging the fact that $\kaplam^{\rm th}$ is relatively insensitive to temperature, 
and recognizing that relaxation times decrease precipitously with $T$ (necessitating longer simulation times),
we used biased simulations at $T = 298$~K to determine the resistance to engulfment, $\glam$, 
using the procedure outlined in section~2D.
We obtained $\kaplam$ by plugging our estimate of $\glam$ in Equation~\ref{eq:kaplam}
along with $\rho = 5 \times 10^4$~mol/m$^3$, $\gamma = 3.7 \times 10^{-5}$~kJ/m$^2$, and $\dm = -0.548$~kJ/mol for TIP4P/Ice at $T = 298$~K~\cite{Abascal2005-tm,Espinosa2016-ko,marks_characterizing_2023}.
As shown in Figures~\ref{fig:wild-sbw}B -- E for sbwAFP, the interfacial curvatures thus calculated, $\kaplam^{\rm th}$, agree well with those estimated using interfacial shape, $\kaplam^{\rm int}$ (using the procedure described in section~5) for each of the four box sizes that we studied.
The maximum curvature, $\kapm^{\rm th}$, was then plotted against $L^{-2}$ (Figure~\ref{fig:wild-sbw}F).

\subsection{Estimating $\Pm$ and $\am$}
\begin{figure}[htbp]
\centering
\includegraphics[width=0.75\textwidth]{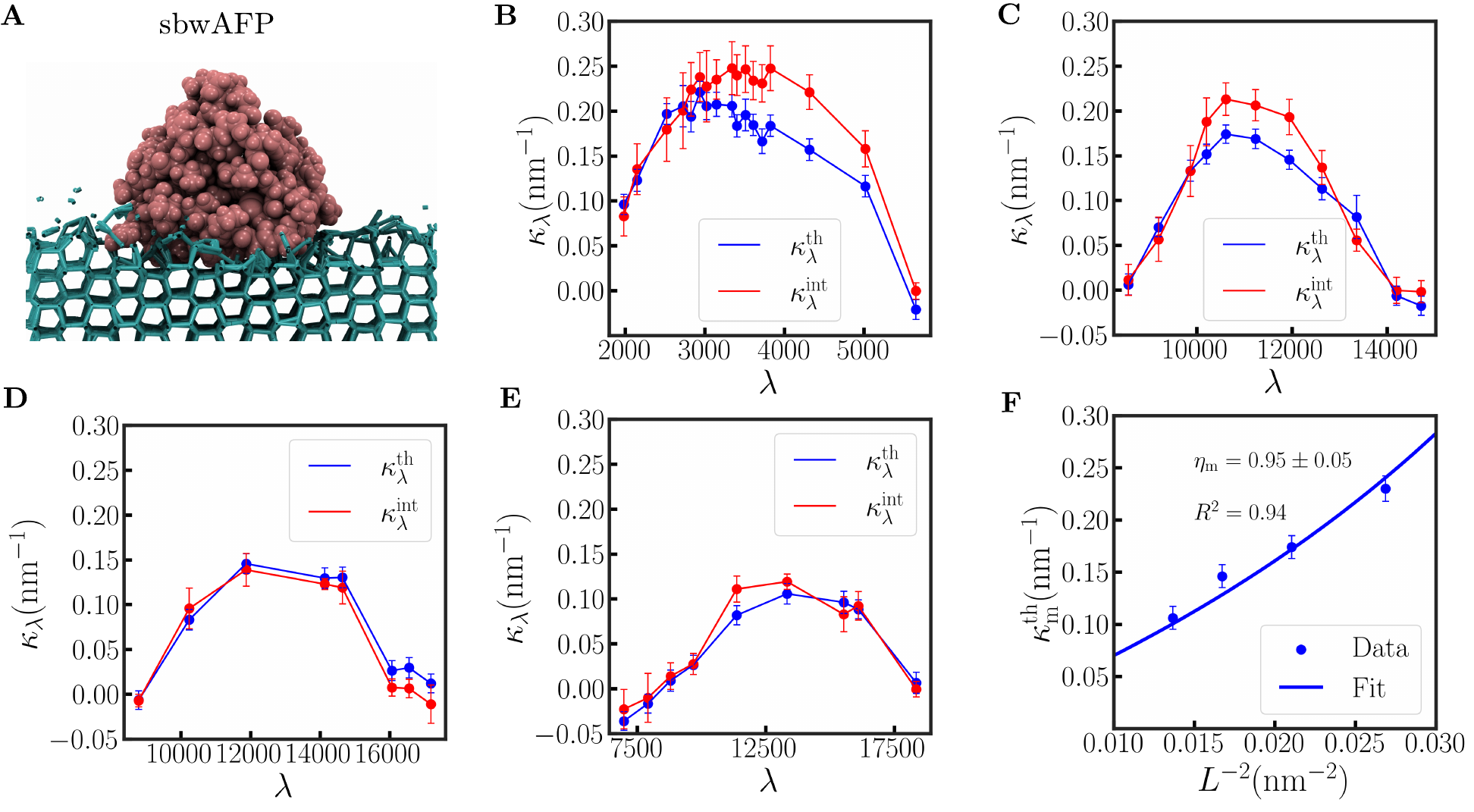}
\caption{
Characterizing the effectiveness of wild-type sbwAFP at resisting engulfment by ice. 
(A) Snapshot of sbwAFP (pink) bound to ice (cyan); liquid water is hidden for clarity.
(B-E) The variation of interfacial curvature, $\kaplam$, with the number of ice-like waters, $\lambda$, is shown for four different simulation box sizes: $5.88 \times 6.33$~nm$^2$ (B), 
$6.62 \times 7.23$~nm$^2$ (C), $7.35 \times 8.13$~nm$^2$ (D), and $8.09 \times 9.04$~nm$^2$ (E). 
Estimates of $\kaplam^{\rm th}$, obtained from theory (Equation~\ref{eq:kaplam}, blue) agree well with those obtained from the shape of the ice-water interface (red).
(F) The maximum interfacial curvature, $\kapm^{\rm th}$, for each simulation box is plotted against $\Lsqinv$ (symbols), and the data is fit to Equation~\ref{eq:si-keta} (line) with $\etam$ as a fit parameter ($\am$ and $\Pm$ were determined from the shape of the AFP).
The simulation data agrees well with Equation~\ref{eq:si-keta} with the best fit being provided by $\etam =0.95$. 
}
\label{fig:wild-sbw}
\end{figure}

To obtain estimates of the perimeter, $\Pm$, of the three-phase contact line corresponding to $\lambda = \lamm$, and the area, $\am$, that it encloses, we used the contact line, $\conl$, determined in section~5C (Figure~\ref{fig:conl}D, red). 
The perimeter, $\Pm$ was computed by adding up the lengths of the line segments that make up the contact line, i.e., $\Pm = \int_{\conl} ds$, whereas the area, $\am$, was calculated using Green's theorem, i.e., $\am = \iint_{\notin \intf} dx dy = 0.5 \int_{\conl}(y dx + x dy)$.
Estimates of $\Pm$ and $\am$ were obtained for each of the four simulation box sizes used to study sbwAFP, and as seen in Table~\ref{tab:pm_am}, they agree well with one another.
To minimize the uncertainty in their estimates, values of $\Pm$ and $\am$ averaged over the different simulation box sizes were used to fit the $\kapm$ vs $L$ data using Equation~\ref{eq:si-keta} (Figure~\ref{fig:wild-sbw}F).
%

\begin{table}[H]
\caption{Estimates of $\Pm$ and $\am$ obtained for the different simulation box dimensions used to study sbwAFP.} 
\centering
\begin{adjustbox}{width=0.35\textwidth}
\small
\begin{tabular}{rlrrrrrrr}
  \hline
 $L_x \times L_y$~(nm$^2$) & $\Pm$~(nm) & $\am$~(nm$^2$)\\ 
  \hline
 5.89 x 6.33 & 13.27 & 11.8 \\
 6.62 x 7.23 & 12.89 & 11.23 \\
 7.35 x 8.13 & 13.5 & 12.03 \\
 8.09 x 9.04 & 12.63 & 10.71 
\end{tabular}
\end{adjustbox}
\label{tab:pm_am}
\end{table} 

%
The best fit to the data was obtained for $\etam = 0.95 \pm 0.05$ suggesting that sbwAFP is optimal at resisting engulfment by ice. 
To estimate the error in $\etam$, we randomly chose a $\kapm$-value for every AFP separation, $L$, 
drawing from a Gaussian distribution with its mean and standard deviation corresponding to $\kapm^{\rm th}$ and its error, respectively.
The set of $\kapm$-choices was then fitted to Equation~\ref{eq:si-keta} to obtain an $\etam$-estimate. 
This procedure was repeated 1000 times and the standard deviation of the $\etam$-estimates was reported as the error in $\etam$.

\section{Data for other AFPs}
\begin{figure}[htbp]
\centering
\includegraphics[width=0.8\textwidth]{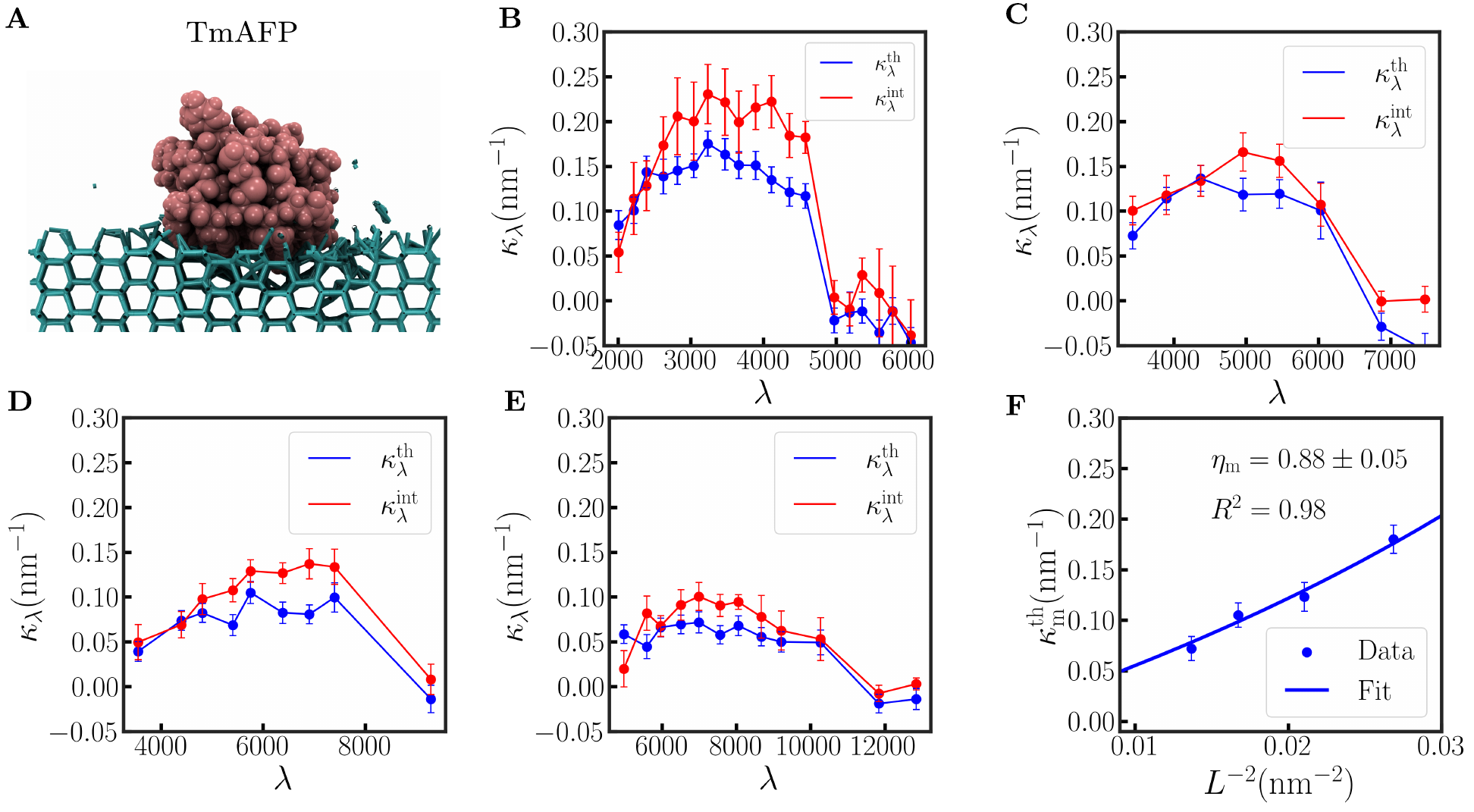}
\caption{
Characterizing the effectiveness of {\it Tm}AFP at resisting engulfment by ice. 
(A) Snapshot of {\it Tm}AFP (pink) bound to ice (cyan); liquid water is hidden for clarity.
(B-E) The variation of interfacial curvature, $\kaplam$, with the number of ice-like waters, $\lambda$, is shown for four different simulation box sizes: $5.88 \times 6.33$~nm$^2$ (B), 
$6.62 \times 7.23$~nm$^2$ (C), $7.35 \times 8.13$~nm$^2$ (D), and $8.09 \times 9.04$~nm$^2$ (E). 
Estimates of $\kaplam^{\rm th}$, obtained from theory (Equation~\ref{eq:kaplam}, blue) agree well with those obtained from the shape of the ice-water interface (red).
(F) The maximum interfacial curvature, $\kapm^{\rm th}$, for each simulation box is plotted against $\Lsqinv$ (symbols), and the data is fit to Equation~\ref{eq:si-keta} (line) with $\etam$ as a fit parameter ($\am$ and $\Pm$ were determined from the shape of the AFP).
The simulation data agrees well with Equation~\ref{eq:si-keta} with the best fit being provided by $\etam =0.88$. 
}
\label{fig:tm}
\end{figure}

\begin{figure}[htbp]
\centering
\includegraphics[width=0.8\textwidth]{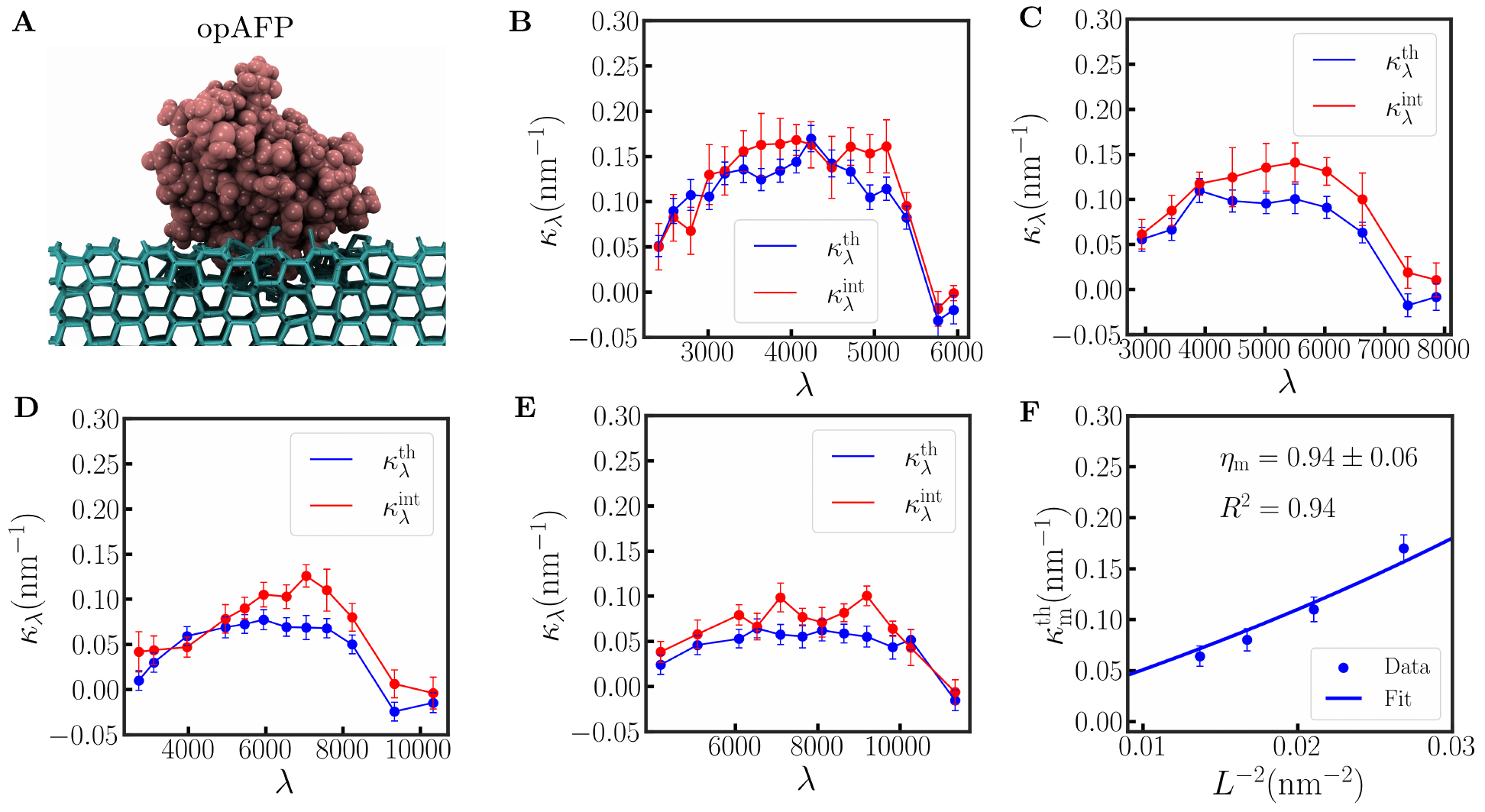}
\caption{
Characterizing the effectiveness of opAFP at resisting engulfment by ice. 
(A) Snapshot of opAFP (pink) bound to ice (cyan); liquid water is hidden for clarity.
(B-E) The variation of interfacial curvature, $\kaplam$, with the number of ice-like waters, $\lambda$, is shown for four different simulation box sizes: $5.88 \times 6.33$~nm$^2$ (B), 
$6.62 \times 7.23$~nm$^2$ (C), $7.35 \times 8.13$~nm$^2$ (D), and $8.09 \times 9.04$~nm$^2$ (E). 
Estimates of $\kaplam^{\rm th}$, obtained from theory (Equation~\ref{eq:kaplam}, blue) agree well with those obtained from the shape of the ice-water interface (red).
(F) The maximum interfacial curvature, $\kapm^{\rm th}$, for each simulation box is plotted against $\Lsqinv$ (symbols), and the data is fit to Equation~\ref{eq:si-keta} (line) with $\etam$ as a fit parameter ($\am$ and $\Pm$ were determined from the shape of the AFP).
The simulation data agrees well with Equation~\ref{eq:si-keta} with the best fit being provided by $\etam =0.94$.
}
\label{fig:op}
\end{figure}

\begin{figure}[htbp]
\centering
\includegraphics[width=0.8\textwidth]{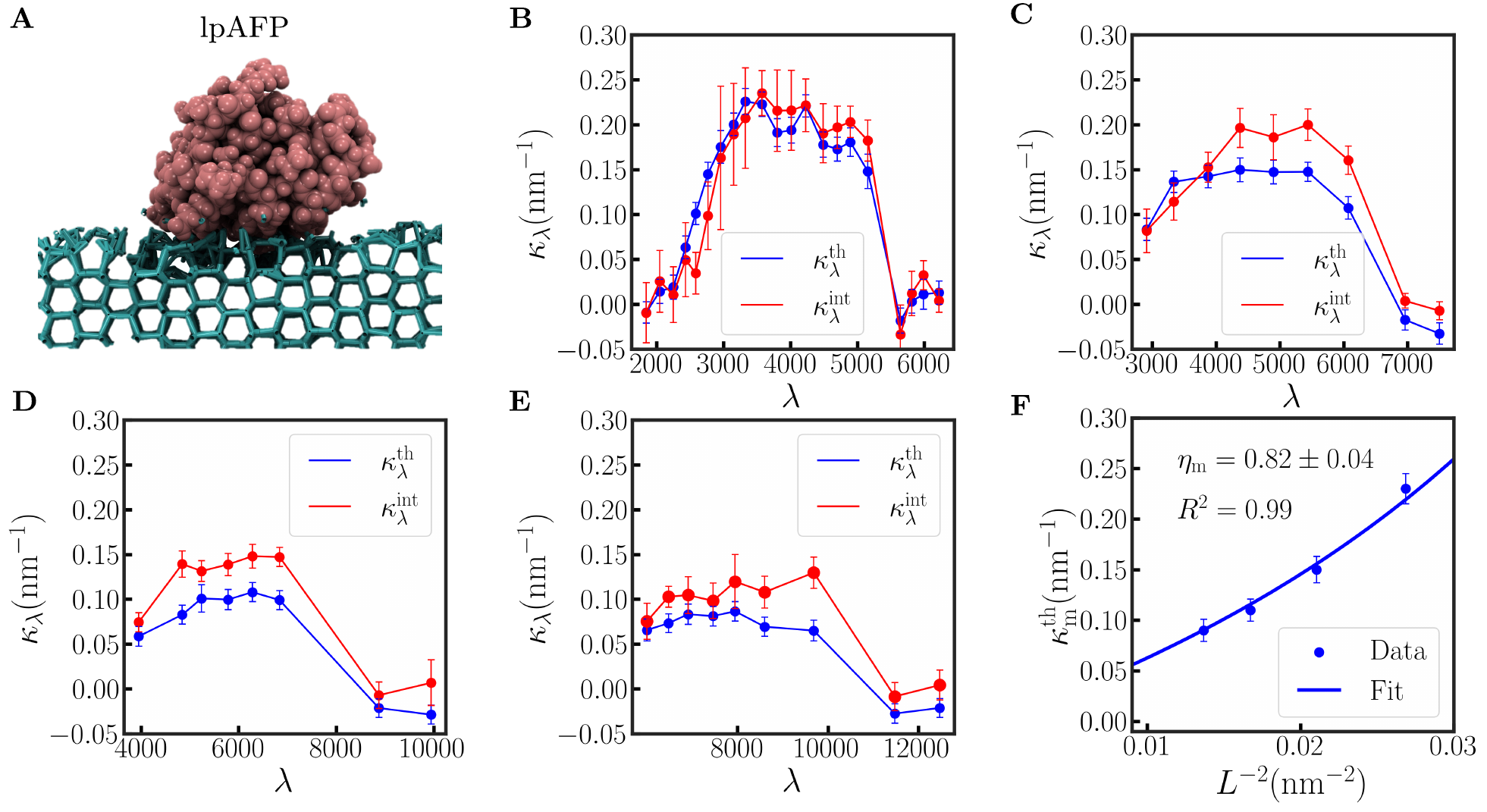}
\caption{
Characterizing the effectiveness of {\it Lp}AFP at resisting engulfment by ice. 
(A) Snapshot of {\it Lp}AFP (pink) bound to ice (cyan); liquid water is hidden for clarity.
(B-E) The variation of interfacial curvature, $\kaplam$, with the number of ice-like waters, $\lambda$, is shown for four different simulation box sizes: $5.88 \times 6.33$~nm$^2$ (B), 
$6.62 \times 7.23$~nm$^2$ (C), $7.35 \times 8.13$~nm$^2$ (D), and $8.09 \times 9.04$~nm$^2$ (E). 
Estimates of $\kaplam^{\rm th}$, obtained from theory (Equation~\ref{eq:kaplam}, blue) agree well with those obtained from the shape of the ice-water interface (red).
(F) The maximum interfacial curvature, $\kapm^{\rm th}$, for each simulation box is plotted against $\Lsqinv$ (symbols), and the data is fit to Equation~\ref{eq:si-keta} (line) with $\etam$ as a fit parameter ($\am$ and $\Pm$ were determined from the shape of the AFP).
The simulation data agrees well with Equation~\ref{eq:si-keta} with the best fit being provided by $\etam =0.82$.
}
\label{fig:lp}
\end{figure}

%
The analysis described in section~7 was repeated for {\it Tm}AFP, opAFP and {\it Lp}AFP, and the corresponding results are shown in Figures~\ref{fig:tm}, \ref{fig:op} and \ref{fig:lp}.

\section{Studying an sbwAFP mutant to interrogate how charged residues on the NBS influence $\etam$}

\subsection{Preparation of mutant sbwAFP structure}
\begin{figure}[htbp]
    \centering
    \includegraphics[width=0.8\linewidth]{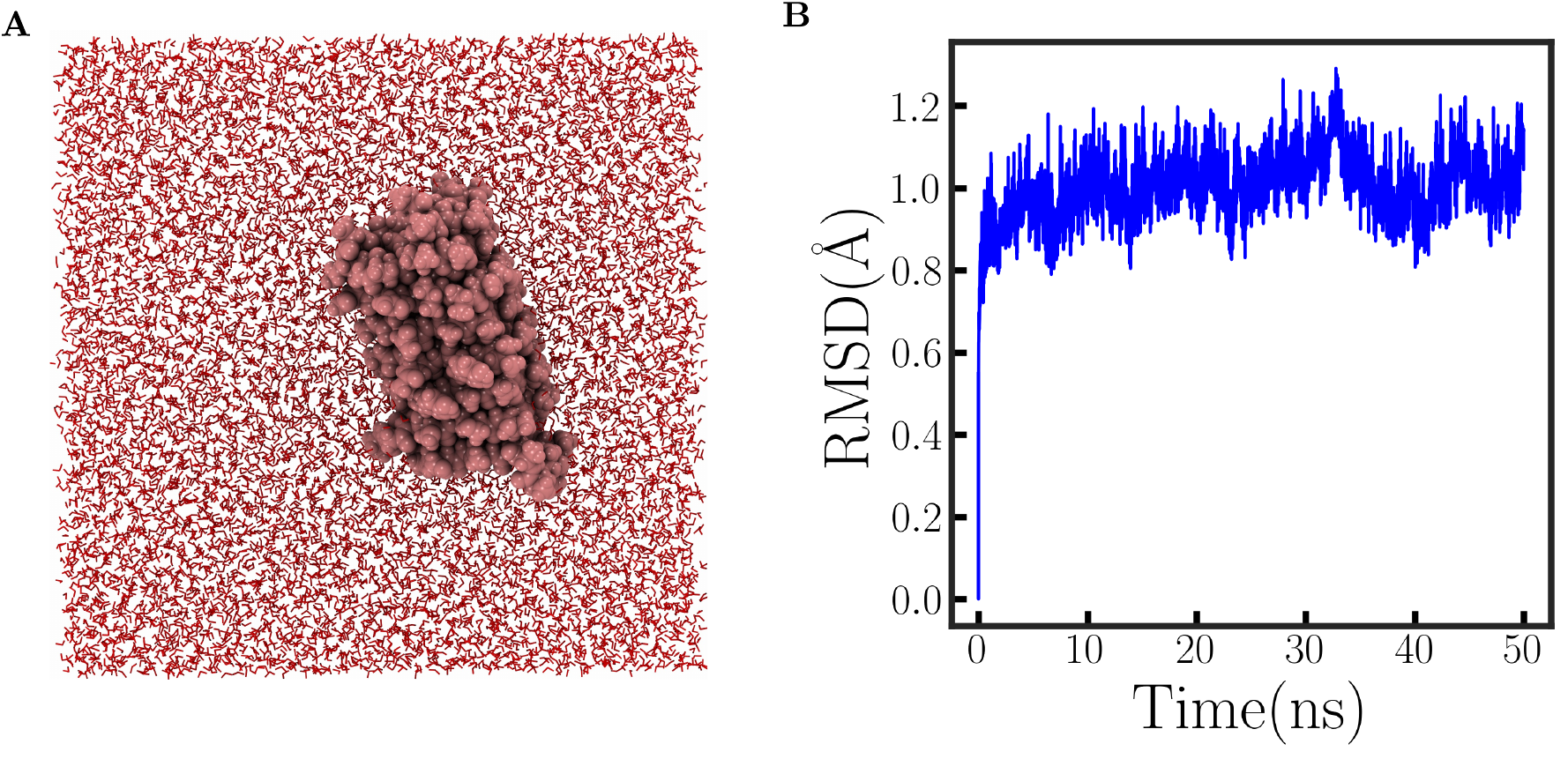}
    \caption{
(A) The sbwAFP mutant retains its folded structure after a 50~ns unbiased simulation. 
(B) The RMSD of protein backbone atoms with respect to the structure at $t=0$ is roughly 0.1~nm, 
highlighting that the folded structure of the mutated protein closely resembles that of the wild-type.
}
    \label{fig:backbone_rmsd}
\end{figure}

The \texttt{swapaa} command in UCSF Chimera \cite{Pettersen2004-ft} was used to make the following mutations from charged residues to alanines in sbwAFP: K19A, D28A, D34A, R40A, D45A, K49A, K50A, D55A, D59A, K60A, R79A, R102A, and K112A. 
To assess the stability of the mutated structure, an unbiased simulation with the hydrated sbwAFP mutant is run at $T = 298$~K for 50~ns, as shown in Figure~\ref{fig:backbone_rmsd}. 
The root mean square deviation (RMSD) of the heavy atoms was computed over the 50~ns simulation trajectory (Figure~\ref{fig:backbone_rmsd}B) and remained below 0.13~nm over the duration of the simulation illustrating the stability of the mutated protein structure.

\subsection{Estimating $\etam$ for the mutant sbwAFP}
\begin{figure}[htbp]
\centering
\includegraphics[width=0.75\textwidth]{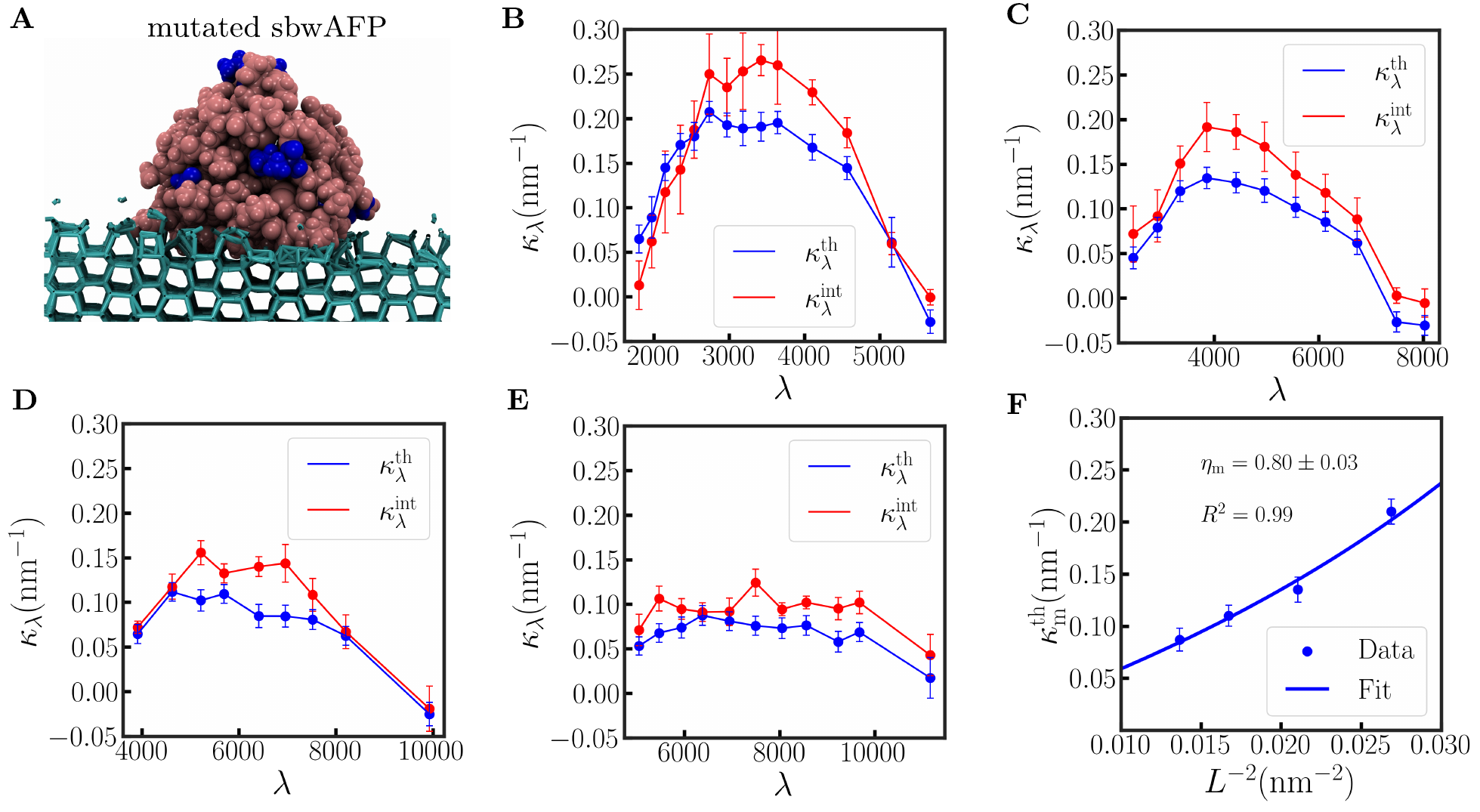}
\caption{
Characterizing the effectiveness of the sbwAFP mutant at resisting engulfment by ice. 
(A) Snapshot of sbwAFP mutant (pink) bound to ice (cyan); the mutated residues are colored blue, and liquid water is hidden for clarity.
(B-E) The variation of interfacial curvature, $\kaplam$, with the number of ice-like waters, $\lambda$, is shown for four different simulation box sizes: $5.88 \times 6.33$~nm$^2$ (B), 
$6.62 \times 7.23$~nm$^2$ (C), $7.35 \times 8.13$~nm$^2$ (D), and $8.09 \times 9.04$~nm$^2$ (E). 
Estimates of $\kaplam^{\rm th}$, obtained from theory (Equation~\ref{eq:kaplam}, blue) agree well with those obtained from the shape of the ice-water interface (red).
(F) The maximum interfacial curvature, $\kapm^{\rm th}$, for each simulation box is plotted against $\Lsqinv$ (symbols), and the data is fit to Equation~\ref{eq:si-keta} (line) with $\etam$ as a fit parameter ($\am$ and $\Pm$ were determined from the shape of the AFP).
The simulation data agrees well with Equation~\ref{eq:si-keta} with the best fit being provided by $\etam =0.8$.
}
\label{fig:mutant-sbw}
\end{figure}

The analysis described in section~7 was repeated for the sbwAFP mutant, 
and the corresponding results are shown in Figure~\ref{fig:mutant-sbw}.

\newpage

\bibliography{supplement}